# Quantum Modelling of Electro-Optic Modulators

José Capmany[1,*] and Carlos R. Fernández-Pousa[2]

[1] *ITEAM Research Institute, Universidad Politécnica de Valencia, 46022 Valencia, Spain*

[2] *Signal Theory and Communications, Dep. of Communications Engineering, Univ. Miguel Hernández, 03202 Elche, Spain*

[*]*Corresponding author: jcapmany@iteam.upv.es*

**Abstract**
Many components that are employed in quantum information and communication systems are well known photonic devices encountered in standard optical fiber communication systems, such as optical beamsplitters, waveguide couplers and junctions, electro-optic modulators and optical fiber links. The use of these photonic devices is becoming increasingly important especially in the context of their possible integration either in a specifically designed system or in an already deployed end-to-end fiber link. Whereas the behavior of these devices is well known under the classical regime, in some cases their operation under quantum conditions is less well understood.

This paper reviews the salient features of the quantum scattering theory describing both the operation of the electro-optic phase and amplitude modulators in discrete and continuous-mode formalisms. This subject is timely and of importance in light of the increasing utilization of these devices in a variety of systems, including quantum key distribution and single-photon wavepacket measurement and conformation. In addition, the paper includes a tutorial development of the use of these models in selected but yet important applications, such as single and multi-tone modulation of photons, two-photon interference with phase-modulated light or the description of amplitude modulation as a quantum operation.

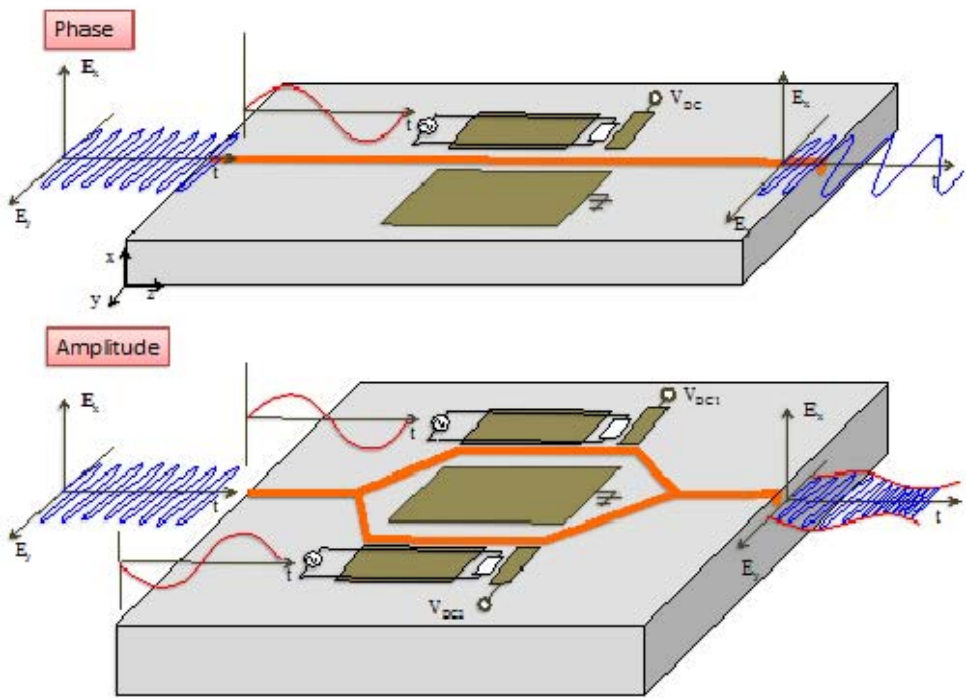

Layout configurations for integrated electro-optic modulators. The upper part shows a phase travelling wave modulator. The lower part shows a Y-Branch travelling wave amplitude modulator.

**Index**

# 1. Introduction

Temporal modulation of light represents one of the key technologies in modern classical optical communication and processing systems. Despite the variety of available modulation devices, electro-optic modulators and, in particular, integrated waveguide modulators, represent one of the most widely employed photonic components [1-5]. The reasons behind their success are various but can be broadly categorized into two groups. On the one hand, they provide unique flexibility as far as the characteristics of the output signal are concerned. Both analog [2] and digital [3], amplitude and phase [4] modulation operations can be easily implemented over a continuous-wave input signal. By means of suitable chip design, carrier suppressed and advanced in-phase-quadrature modulation formats can also be implemented [6-8]. Furthermore, chirped (positive and negative) and chirpless output modulated signals can also be produced by suitable control of their biasing and modulation voltage waveforms [9]. The second reason for their widespread use is connected to their broadband operation. Commercially available electro-optic modulators currently provide modulation bandwidths of 50 GHz [10] and several experimental demonstrations have reported bandwidths in excess of 100 GHz [11]. This is possible since highly sophisticated transmission line designs for the modulation electrodes can be incorporated into the chip design of the device, allowing for an extremely efficiently propagation speed matching between the optical and electrical signals [12]. In this paper we mainly concentrate on these integrated waveguide configurations, although the results, models and operating principles described here can be readily extended to bulk-optics modulators.

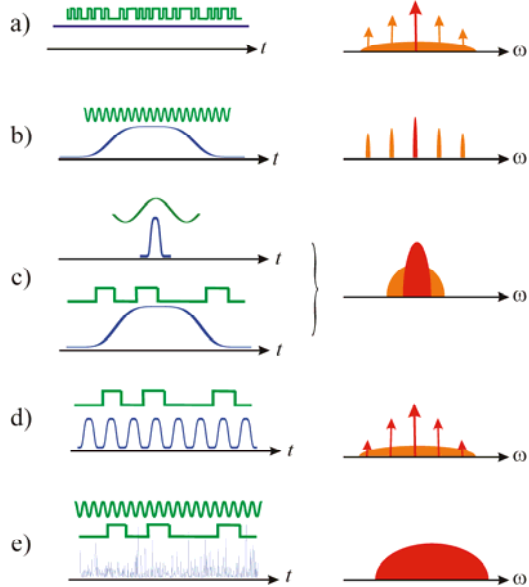

**Figure 1.1.** Operation regimes of optical modulators. Left: intensity of lightwave signals (blue) and modulation waveforms (green). Right: optical spectrum of the input signals (red) and modulated signals (orange).

From the system's point of view, electro-optic modulators implement at classical level a linear multiplicative transformation of the complex envelope $E(t)$ describing the temporal profile of radiation,

$$E'(t) = m(t)E(t) \qquad (1.1)$$

where $m(t)$ is the phase or amplitude modulating function. This functions depends on an external, user-defined voltage driving signal $V(t)$ whose bandwidth may reach the microwave range as mentioned above. Then, several operation regimes can be recognized depending on the relative bandwidth of $E(t)$ and $m(t)$. Indeed, Eq. (1.1) implies that the spectral width of the output envelope $B_{E'}$ equals the sum of the spectral width of the input radiation, $B_E$, plus the bandwidth of the modulation function, $B_m$. Referring to figure 1.1, a first regime is associated to the limit where incoming radiation is narrowband with respect to modulation bandwidth, $B_E<<B_m$. In this case, modulators enlarge the spectral width of the input radiation, which can be continuous-wave (cw) or pulsed, resulting in a broad continuous spectrum if $m(t)$ is random (figure 1.1a) or in the generation of spectral sidebands if $m(t)$ is periodic (figure 1.1b). This regime is typical of analog or digital transmission systems where narrowband cw light, typically a laser, is modulated by the data-bearing function $m(t)$. A second possibility, shown in figure 1.1c, is that both bandwidths are similar, $B_E \sim B_m$, implying a moderate spectral increase. For pulsed input signals, for instance, this regime finds applications in pulse processing systems. Figures 1.1d-e exemplify the third possibility, $B_E >> B_m$. In the first case, figure 1.1d, the input is pulsed and modulation is slow compared to pulse duration or, as a limit case, constant. The output pulse's spectral width is not enlarged and $m(t)$ simply controls its global phase and amplitude. Digital transmission systems, where binary information in encoded in either the phase or amplitude of a pulse sequence, belong to this category. Finally, in figure 1.1e we illustrate the modulation of broadband, continuous-wave chaotic light, such as that of a LED or an ASE source, a possibility that finds application in a number of photonic processing systems.

The versatility of these devices is such that their use has extended to other fields, in particular to quantum information and quantum communication systems. Modulators are key devices for quantum control, especially in feed-forward systems, with a wealth of applications such as squeezing control, quantum teleportation, quantum cloning, quantum non-demolition measurements or quantum computation based on cluster states (see [13, 14] and references therein). In quantum key distribution (QKD) [15, 16], phase and amplitude modulation of weak coherent pulses as sources of approximated single-photon states are used in phase-coded (PC) QKD systems [17-22], and also in frequency-coded (FC) [23-29] and subcarrier multiplexed (SCM) [30, 31] QKD systems. They are also used for controlling the amplitude of coherent states in Continuous-Variable QKD systems, either with a gaussian distribution [32-39] or with discrete modulation [40], and also in distributed-phase-reference protocols [41-46].

Modulators also permit the extension to the photon level of standard techniques of classical photonic signal processing such as temporal imaging [47]. As for photon wavepacket conformation, amplitude and phase modulators have been used for tailoring the wave function of heralded single photons in both phase [48] and amplitude [49], and modulation-based measurement techniques of biphoton wavefunctions have been demonstrated [50]. At a more fundamental level, phase modulation of entangled photon pairs has been shown [51, 52] to produce non-local effects similar to dispersion [53], and the high-dimensional frequency correlations introduced by phase modulation in spontaneous parametric down-converted light have been recently studied [54]. Finally, synchronous driving fields with pseudo-random bit sequences permit enlarging the single-photon bandwidth and subsequently recovering the wavepacket, thus augmenting its resilience to both noise and potential narrowband jammers in analogy to the spread-spectrum technique used in radiocommunications [55]. In summary, and due to their inherent capabilities for modifying the spectral content of radiation together with its ease of integration, it is envisaged that the use of modulators will expand to other emerging applications in the field of quantum information and processing systems [56].

In this context, the development and practical implementation of quantum systems requires, at a certain step, not only a knowledge of the underlying physical principles, but also a considerable engineering design step for which models of the essential photonic components are required. While it is true that these models are available for classical passive optical devices, where the absence of coupling between different frequencies permits a description based on a single frequency mode [57-61], it has only been recently that the study of electro-optic modulators as multi-mode quantum scattering devices has gained some attention, both as far as phase [62-64] and amplitude [65] modulation are concerned. This can in part be due to the different approaches used in the analysis of specific systems. In QKD systems, for instance, where weak coherent pulses are used together with sinusoidal driving voltages, the discrete-mode formalism is sufficient. In turn, the combination of arbitrary driving voltages and/or wideband photon wavepackets or, in general, wideband radiation, inevitably requires the continuous-mode formalism. A theory of electro-optic modulators should be sufficiently versatile to incorporate these different views in a unified framework, and also allow for the introduction of simplifying approximations in concrete applications.

The aim of this paper is double. On the one hand, in sections 2, 3 and 4 it is reviewed the salient features of the scattering theory describing both the operation of the electro-optic phase (section 3) and amplitude (section 4) modulators subject to single photon and coherent input states. This review incorporates, as an added value, an extension of the results from the discrete mode to the continuous mode formalism. The second purpose of this paper is to provide a tutorial development of the use of these models in selected but yet important applications. Several illustrative examples have been selected, which are developed in section 5 and cover a wide range of topics. Thus, in subsections 5.2 and 5.3 the modulation by single and multiple analog sinusoids of a single photon input state are analyzed. This analysis is of interest for FC QKD systems, which are studied with more detail in subsection 5.4. Subsection 5.5 treats the modulation of two photons (one per port) input states and the possibility of switching between entangled and separable output states. Single-tone modulation of single photon wavepackets is developed in subsection 5.6, while two-photon interference with phase-modulated wavepackets is treated in 5.7. The general consideration of amplitude electro-optic modulation of single photons as a quantum operation is addressed in subsection 5.8, and the section is closed in 5.9 with a description of the correlations of phase and amplitude modulated states. Finally, section 6 summarizes and closes the paper.

## 2. Quantum models for passive optical splitters

*2.1 Bulk-optics beamsplitter*

The operation of the bulk optics beamsplitter (BS) under quantum regime can be found in several texts and references in the literature [57-61], where excellent descriptions based on the Heisenberg picture are developed. The presentation here will be based on the Schrödinger picture that, for reasons to be apparent later in this paper, is more convenient to the present purposes. The description is then specialised to the two more prominent guided-wave versions of the device [4, 66, 67]; the Directional Coupler (DC) and the Y-Branch power splitter (YB), which are employed in the electro-optic amplitude modulator (EOM). To describe the quantum operation of the optical beamsplitter the reader is referred to figure 2.1. The upper part of the figure describes the general framework of the Hilbert spaces and states that characterize both the input and the output of the BS. It is assumed that the input is given by the product state $|in\rangle$ =$|\Psi_1\rangle_1 \otimes |\Psi_2\rangle_2$, where $|\Psi_j\rangle_j$ belongs to $H_j$, the multimode Hilbert space corresponding to input port $j$=1,2. Although the Hilbert space is assumed multimode, the BS is a single-mode device and does not mix frequencies, so it is enough to analyse its operation over a unique mode.

In the same way, an output state from the BS is given by $|out\rangle = |\Phi_{1'}\rangle_{1'} \otimes |\Phi_{2'}\rangle_{2'}$, with $|\Phi_{j'}\rangle_{j'} \in H_{j'}$ the Hilbert space of the states of output port $j'$=1′,2′. Upon these definitions, the action of the BS in the Schrödinger picture is described by a unitary scattering operator acting on input states $\hat{S}_{BS}|in\rangle=|out\rangle$. It leaves invariant the two-port vacuum state, $\hat{S}_{BS}|vac\rangle \otimes |vac\rangle = |vac\rangle \otimes |vac\rangle$ and its action over an arbitrary Fock state can be reduced to its action over creation operators. With the usual notations for the creation operators $\hat{a}_1^+ = \hat{a}^+ \otimes \hat{1}$ and $\hat{a}_2^+ = \hat{1} \otimes \hat{a}^+$ in the input ports it can be expressed in matrix form:

$$\hat{S}_{BS} \begin{pmatrix} \hat{a}_1^+ \\ \hat{a}_2^+ \end{pmatrix} \hat{S}_{BS}^+ = \begin{pmatrix} t' & r' \\ r & t \end{pmatrix} \cdot \begin{pmatrix} \hat{a}_1^+ \\ \hat{a}_2^+ \end{pmatrix} \qquad (2.1)$$

Here, $t'$ and $r'$ define the field transmission and reflection coefficients for a classical signal fed through port 1 and correspondingly $t$ and $r$ are those for inputs from port 2, as shown in figure 2.1. For instance, for a one-photon input in the first port one has

$$\begin{aligned}\hat{S}_{BS}|1\rangle \otimes |vac\rangle &= \hat{S}_{BS}(\hat{a}^+ \otimes \hat{1})|vac\rangle \otimes |vac\rangle \\ &= \hat{S}_{BS}(\hat{a}^+ \otimes \hat{1})\hat{S}_{BS}^+|vac\rangle \otimes |vac\rangle = t'|1\rangle \otimes |vac\rangle + r'|vac\rangle \otimes |1\rangle\end{aligned} \qquad (2.2)$$

The unitarity of the matrix in Eq. (2.1) leads to the energy conservation and the reciprocity relationships, which are:

$$|t'|^2 + |r'|^2 = |t|^2 + |r|^2 = 1 \qquad r^*t' + r't^* = r^*t + r't^* = 0 \qquad (2.3)$$

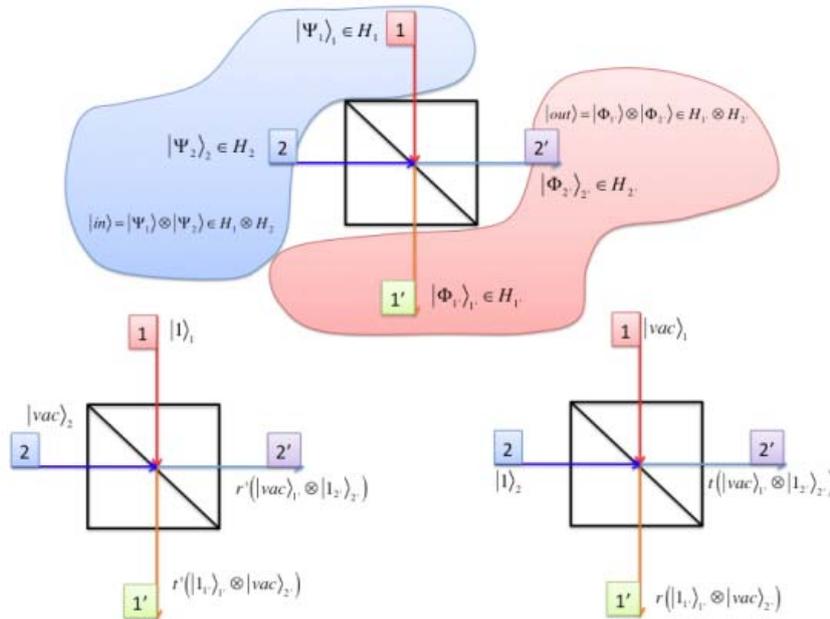

**Figure 2.1:** Quantum state labeling (upper) and single photon behavior (lower) of an optical beamsplitter

*2.2 Directional coupler*

The most widespread version of the optical beamsplitter in guided-wave format is the directional coupler (DC) or 2×2 coupler [4, 66-68]. It is composed of two input and two output optical fibers or integrated input waveguides. Signal coupling is achieved on the central part of the device by creating the adequate conditions such that the evanescent fields of the fundamental guided mode in one waveguide can excite the fundamental guided mode in the other and vice-versa. Different techniques to achieve signal coupling and for analyzing the device performance and carry out a proper design have been developed in the last 20 years and are quite well known and understood [4]. The beamsplitting power ratio of the 2×2 coupler is characterized by its coupling constant $k$ [4] that defines the fraction of power coupling (i.e., crossing) from one waveguide to the other. Another characteristic is the $\pi/2$ phase shift that the optical field experiences when coupling from one waveguide to the other, which means that the reflection coefficients $r$, $r'$ in Eq. (2.1) are imaginary. The results for the general bulk-optics beamsplitter can thus be employed to describe the 2×2 directional coupler by making:

$$t = t' = \sqrt{1-k} \qquad r = r' = i\sqrt{k} \qquad (2.4)$$

*2.3 Y-branch power splitter*

The Y-Branch or 1×2 Power Splitter is another popular guided-wave implementation of the optical beamsplitter although it is more commonly employed in its integrated optics version than in optical fiber format [67]. As the 2×2 directional coupler it is characterized by its coupling constant $k$ (usually $k=1/2$) that again defines the fraction of power coupling or crossing from one waveguide to the other. However, even under classical regime, the Y-Branch operation has some distinctive features that are sometimes misunderstood. To clarify its operation a brief description of the device operation is now provided for the case of $k=1/2$. To do this, consider the field modal structure at the device input and output sides. The Y branch can operate as a power splitter or as a power combiner, as shown respectively in the upper and lower parts of figure 2.2.

When it operates as a power splitter the field structure at the input port $E_{in}(x,y)$ is proportional to that of the fundamental mode of the input waveguide, that is $E_{in}(x,y) \propto E_0(x,y)$. At the output side (right) the device is composed of two separate waveguides separated by a distance $S$. Defining the transversal model field variations in the upper and lower waveguides by $E_{DS}(x,y)$ and $E_{DI}(x,y)$, respectively:

$$E_{DS}(x,y) = E_0(x, y - S/2) \qquad E_{DI}(x,y) = E_0(x, y + S/2) \qquad (2.5)$$

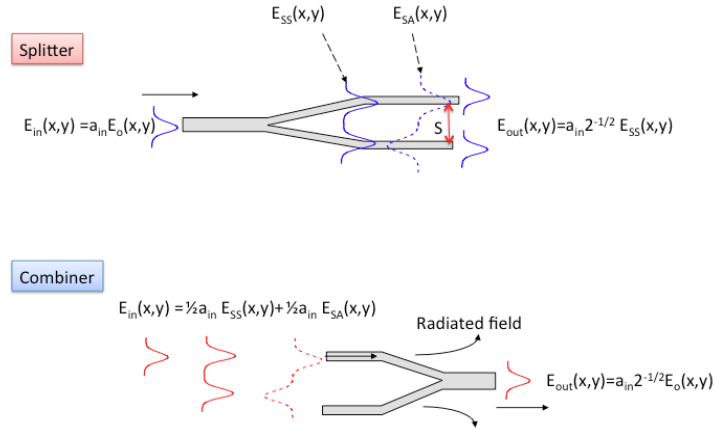

**Figure 2.2.** Classical behaviour of Y-branch guided-wave even ($k=1/2$) power splitter (upper) and combiner (lower)

It can as well be considered that the behavior of the structure at the output is defined by its own compound modes or supermodes which result from the linear combination of $E_{DS}(x,y)$ and $E_{DI}(x,y)$. This combination can be in the form of an addition yielding a symmetric supermode $E_{SS}(x,y)$, or in the form of a subtraction yielding an asymmetric supermode $E_{SA}(x,y)$. Hence:

$$E_{SS}(x,y) = E_{DS}(x,y) + E_{DI}(x,y) \qquad E_{SA}(x,y) = E_{DS}(x,y) - E_{DI}(x,y) \qquad (2.6)$$

Since these supermodes are those that are characteristic of the composite structure, it follows that any field propagating through it can be expressed as a linear combination of them. For instance, if the field injected at the device is given by:

$$E_{in}(x,y) = a_{in} E_0(x,y) \qquad (2.7)$$

where $a_{in}$ represents the mode amplitude, the field structure at the input waveguide is symmetric and upon arriving at the transition zone between the input and output waveguides tends to transform into a symmetric field distribution, with respect to the line dividing the output part of the device into two equal regions. In other words, the input field will only excite the symmetric supermode in the output part of the device. Hence,

$$E_{out}(x,y) = a_{out} E_{SS}(x,y) = a_{out} \left[ E_{DS}(x,y) + E_{DI}(x,y) \right] \quad (2.8)$$

Assuming a lossless device, energy conservation requires that $\int dxdy |E_{in}(x,y)|^2 = \int dxdy |E_{out}(x,y)|^2$. Taking Eq. (2.8) into account and considering that there is no spatial overlapping between the fields in the output waveguides of the fields ($E_{DS}(x,y) \times E_{DI}(x,y)^* = 0$), one has:

$$a_{out} = \frac{a_{in}}{\sqrt{2}} \Rightarrow E_s(x,y) = \frac{a_{in}}{\sqrt{2}} E_{SS}(x) = \frac{a_{in}}{\sqrt{2}} \left[ E_{DS}(x,y) + E_{DI}(x,y) \right] \quad (2.9)$$

and therefore the optical power at the device input is evenly divided into its two output waveguides thus confirming the operation of the Y-Branch as a power divider. Note that there is no extra optical phase shift, and the two modes $E_{DS}$, $E_{DI}$ exit in phase.

Let us now describe the device operation as a signal combiner. Consider for instance the case when the input signal is present only in one waveguide (if the two waveguides are excited then the superposition principle can be applied). The input field can be expressed as the sum or difference of the two supermodes each carrying half the input power:

$$E_{in}(x,y) = a'_{in} E_0(x, y \mp S/2) = \frac{a'_{in}}{2} \left[ E_{SS}(x,y) \pm E_{SA}(x,y) \right] \quad (2.10)$$

with the upper (lower) sign corresponding to the case where the field is injected through the upper (lower) port. When this field arrives to the transition zone it must excite the fundamental (symmetric) mode of the output waveguide. This implies that the symmetric supermode is gradually transformed in this region to the guided output mode $E_0(x,y)$. In turn, the antisymmetric supermode does not couple to the output waveguide and is *radiated* to the exterior part of the device as a certain radiated wave $\pm E_R(x,y)$. The output field in the waveguide is then of the form:

$$E_{out}(x,y) = a'_{out} E_0(x,y) \quad (2.11)$$

and only carries half of the input power, so that:

$$a'_{out} = \frac{a'_{in}}{\sqrt{2}} \quad (2.12)$$

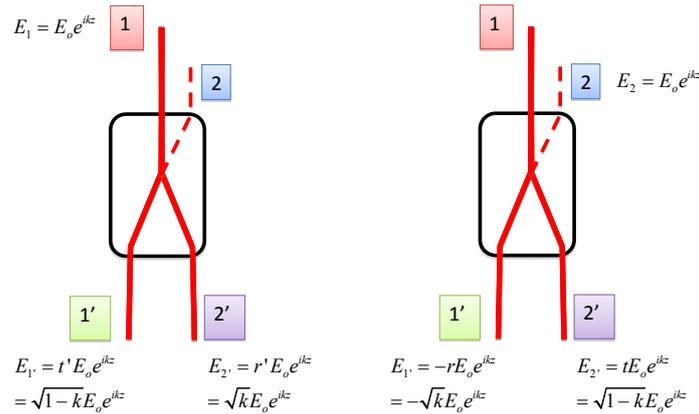

**Figure 2.3.** Classical behaviour of Y-branch guided-wave power splitter using a four port (three real, one virtual) layout.

The consequence is that since in this case the Y-Branch has one physical output port, there is the need to assume that there is a second virtual output port through which the radiated field excited by the antisymmetric supermode is radiated to the exterior. On the reverse operation (power splitter) this port has no physical access from the classical point of view. This observation is fundamental since it should be borne in mind that under quantum regime this port is in the vacuum state [59]. Thus, to correctly represent the Y-Branch beam splitter the layout represented in figure 2.3 will be considered, where the additional radiative port is depicted with a broken line and characterize it in terms of a parameter $k$ that accounts for the fraction of power coupling (i.e crossing) from one waveguide to the other. Then, Eq. (2.1) takes the form:

$$\hat{S}_{YB}\begin{pmatrix}\hat{a}_1^+\\\hat{a}_2^+\end{pmatrix}\hat{S}_{YB}^+ = \begin{pmatrix}\sqrt{1-k} & \sqrt{k}\\-\sqrt{k} & \sqrt{1-k}\end{pmatrix}\cdot\begin{pmatrix}\hat{a}_1^+\\\hat{a}_2^+\end{pmatrix} \quad (2.13)$$

and for the Y-Branch acting as a beam combiner, the matrix is the transpose of that in Eq. (2.13):

## 3. Quantum model of the Electro-optic Phase Modulator

*3.1 Introductory remarks*

In this section quantum models for the scattering theory describing the action of electro-optic phase modulators are developed, with different degrees of complexity and applicability. As a starting point a short account of the classical models of these devices [4,5] is provided, followed by a description on the problems that may show up from a direct translation of this classical description to the quantum level. Figure 3.1 shows a typical configuration of an integrated electro-optic phase modulator [68]. As it can be observed it is composed of a waveguide which is formed by diffusing an electro-optic material (typically Ti) into a depth $d$ (typically around 5-10 μm) of a dielectric substrate (typically $LiNbO_3$) and a set of driving electrodes which span along a distance $l$ (typically between 2-5 cm) along a given propagation axis (for simplicity we consider the $z$ axis and the region between $z = 0$ and $z = l$). The electrodes are divided into two parts. One part has a lumped structure and is employed to provide a dc or bias voltage $V_{DC}$. The second part is implemented by means of a load-matched transmission line designed to propagate a time varying (modulating) voltage $\Delta V(t)$ with a group velocity as close as possible to that of the optical wave inside the optical waveguide. This group velocity matching is essential in order to obtain very high modulation bandwidths (in practice over 20 GHz are routinely obtained in commercial devices). The electro-optic material is such that it is able to linearly change its static refractive index $\eta_0$ along a given cartesian direction in response to an applied voltage $V$ to the electrodes. In other words, $\eta(V) = \eta_0 + (r\eta_0^3 V/2d)$, where $r$ represents the electro-optic coefficient in that particular direction. Usually the values of $r$ depend on the chosen material, the optical wavelength and change from one direction to other. For example, for $LiNbO_3$ $r$=5-18 *pm/V* [4, 68]. The consequence for phase modulators is that for optimum operation the state of polarization of the input field to the waveguide must be linear and aligned with one of the cartesian axes of the transverse plane to the direction of propagation (in the case of figure 3.1 it is shown aligned with the $y$ axis). Although Ti diffused $LiNbO_3$ waveguide electro-optic modulators are the most common devices commercially available, electro-optic modulators can be also implemented using semiconductor materials such as AsGa or polymers. Electro-optic phase modulators can also be implemented using bulk optics configurations as described in [4, 68]. The main advantage of these configurations resides in their lower insertion losses (0.1-0.2 dB) as compared to integrated configurations (0.5-1 dB) since antireflection dielectric stack coatings can be grown at the input and output surfaces. Their main disadvantage stems from their reduced bandwidth (from dc to around 500 MHz is the best reported figure) although bulk optics modulators can be also operated at microwave frequencies (>1GHz) by means of a resonant biasing circuit.

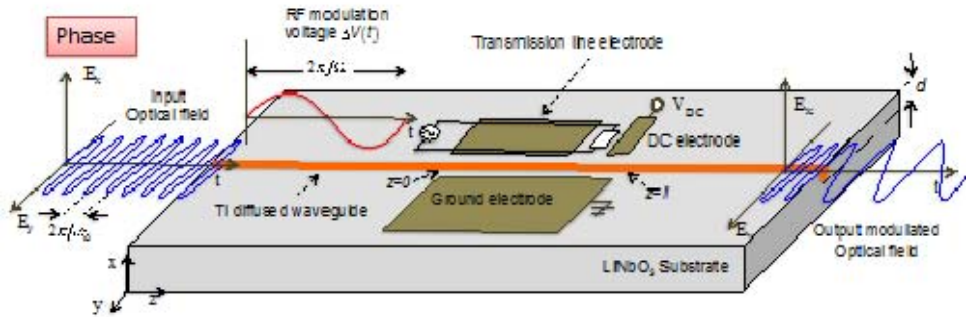

**Figure 3.1.** Layout of an electro-optic integrated phase modulator.

Let us consider the situation depicted in figure 3.1, where a linearly polarized ($y$ direction) monochromatic wave at frequency $\omega_0$ propagates in the positive $z$ direction of a dielectric material with static refractive index $\eta_0$,

$$E_{in}(z,t) = E_0 \exp[ik(\omega_0)z - i\omega_0 t] \quad (3.1)$$

where $k(\omega_0) = \omega_0 \eta_0/c$ is the propagation constant, $c$ is the speed of light in vacuum and $E_0$ its amplitude. This wave enters the transverse phase modulator of thickness $d$ and length $l$ located, as explained before, between $z = 0$ and $z = l$. The modulator is driven by a voltage function $V(t) = V_{DC} + \Delta V(t)$. At the exit plane, the wave is

$$E_{out}(l,t) = E_0 e^{-i\omega_0 t} e^{i\varphi_B} \exp(-i\pi \Delta V(t)/V_\pi) = E_{in}(0,t) e^{i\varphi_B} \exp(-i\pi \Delta V(t)/V_\pi) \quad (3.2)$$

where the constant, bias-dependent phase

$$\varphi_B = \frac{\omega_0}{c}\left(\eta_0 l - \frac{\eta_0^3}{2} r V_{DC} \frac{l}{d}\right) = k(\omega_0) l - \pi \frac{V_{DC}}{V_\pi} \quad (3.3)$$

accounts for the propagation phase shift and the phase shift due to the presence of a bias voltage in a material with electro-optic coefficient $r$, and $V_\pi = \lambda_0 d/\eta_0^3 r l$ is the modulator's half-wave voltage, which is defined as the required dc voltage value to provide a $\pi$ phase shift of the input optical signal, as it is immediately observed from Eqs. (3.2) and (3.3) if $\Delta V(t)=0$. Typical values of $V_\pi$ are in the 2-5 volt range for integrated devices and at least one order of magnitude larger for bulk-optics modulators.

If the input wave is modulated, phase modulation is described in terms of slowly-varying complex envelopes $A_{in}(t)$ and $A_{out}(t)$, defined as

$$E_{in}(0,t) = A_{in}(t) e^{-i\omega_0 t} \qquad E_{out}(l,t) = A_{out}(t) e^{-i\omega_0 t} \quad (3.4)$$

so that, assuming absence of dispersion inside the modulator, the exit envelope is:

$$A_{out}(t) = A_{in}(t) e^{i\varphi_B} \exp(-i\pi \Delta V(t)/V_\pi) \quad (3.5)$$

and therefore the action of the phase modulator always leads to an increase of optical bandwidth. For the particular case of monochromatic inputs, this is viewed as a process of sideband generation. Setting $A_{in}(t) = E_0$ and restricting to a sinusoidal modulation of frequency $\Omega_0$ given by:

$$\Delta V(t) = V_m \cos(\Omega_0 t + \theta) \quad (3.6)$$

getting:

$$A_{out}(t) = E_0 e^{i\varphi_B} \exp[-im\cos(\Omega_0 t + \theta)] = E_0 \sum_{n=-\infty}^{\infty} C_n e^{-in\Omega_0 t} \quad (3.7)$$

where $m = \pi V_m/V_\pi$ is the modulation index, the coefficients are

$$C_n = e^{i\varphi_B}(-ie^{-i\theta})^n J_n(m) \quad (3.8)$$

and where the Jacobi-Anger expansion in terms of the Bessel functions of first kind has been employed:

$$e^{-iz\cos\vartheta} = \sum_{n=-\infty}^{\infty} (-i)^n J_n(z) e^{-in\vartheta} \quad (3.9)$$

The construction of a quantum scattering theory of phase modulation can start in principle from Eq. (3.7) for monochromatic inputs and single-tone driving fields, or from Eq. (3.5) for general input and driving fields. However, some effects should be analyzed with more detail. First, dispersion tends to mismatch the phase of the waves inside the modulator and results in a decrease of modulation bandwidth. In the classical description, Eq. (3.7), this effect may be taken into account by considering that the modulation index $m$ attained by the modulator at a certain voltage level depends on the modulating radio-frequency tone $\Omega_0$, a procedure that will be followed below.

A second effect is related to the observation that modulators are guided media only in a certain frequency range in the optical band. In principle, it could be attempted a quantum description restricted to this band, which is precisely the range of validity where classical formulas such as in Eqs. (3.5) and (3.7) are meaningful. However, it is difficult to substantiate such an observation since the effect of phase modulation is precisely to increase the optical bandwidth: once the input set of frequencies is established, the output set is inevitably enlarged. This leads to consider a description where all input frequencies, even those which are not optical, are allowed, and where expressions such as Eqs. (3.5) or (3.7) are merely effective descriptions which can be recovered when the inputs are optical. In this spirit, and leaving aside the input or output spatial distribution of the frequency modes, the process of sideband generation in Eq.

(3.7), for instance, can still be posed as the determination of input-output transition amplitudes in a scatterer that tends to increase or decrease the input mode frequency by multiples of an amount $\Omega_0$. An expression such as Eq. (3.5) could then be recovered by an appropriate continuous-mode limit.

There is, finally, a more fundamental issue related to unitarity. In the description given by Eqs. (3.5) or (3.7) phase modulation of a carrier will always develop unphysical negative-frequency modes, even classically. For instance, starting form Eq. (3.7) and going back to the electric field at the modulator's output ($z > 0$) leads to:

$$E_{out}(z,t) = E_0 \sum_{n=-\infty}^{\infty} C_n \exp[ik(\omega_n)z - i\omega_n t] \tag{3.10}$$

where $\omega_n = \omega_0 + n\Omega_0$ is the optical frequency of the $n$-th sideband. It can be observed that the infinite number of terms in expansion (3.9) leads to unphysical negative frequencies for sufficiently large values of index $n$. In the classical description of phase modulation this is not a severe problem: under practical operation conditions, $\omega_0/2\pi \sim 200$ THz and $\Omega_0/2\pi < 50$ GHz, so that the ratio $\omega_0/\Omega$ representing the number of necessary sidebands for generating spurious negative-frequency modes is $\approx 10^4$ for inputs with optical frequencies. In addition, and according to Carson's rule [69], the amplitude of the generated sidebands $C_n \sim J_n(m)$ can be neglected for $|n| > m + 1$. Since practical modulation indices $m$ never exceeds the order of ten, this implies that the possibility of generating negative frequencies is of little concern. In a quantum theory, however, the interpretation of these non-positive frequencies is to be analyzed in detail because all modes, even those with negligible probability of generation, contribute to unitarity at the same footing.

As shown below, a truly unitary quantum scattering operator can be constructed and solved in the simplest situation, which is that of a single-tone driven phase modulation [63,64]. The quantum analogue of Eq. (3.7) is then recovered as an approximation valid for inputs belonging to optical bands, where unitarity requirements associated to the existence of a lower bound in frequency are negligible. This route provides a sound footing to several theories of quantum phase modulation based on the sideband generation picture [29, 25, 62]. From that, a general phase modulation operator for arbitrary driving function can be defined. In that case the theory can only be solved perturbatively, and expressions such as Eq. (3.5) are recovered again as an effective description when input states are optical, as also found in theories employing phase modulation of photon wavepackets and/or general driving signals [47, 49].

### 3.2 Single-tone driven phase modulation in discrete-mode formalism

The discrete character of the output frequencies observed under single-mode inputs at frequency $\omega_0$ and single-tone driving justifies a description where only modes with frequencies $\omega_0 + n\Omega_0$ are taken into account, with the only requirement that $n > -\omega_0/\Omega_0$ to avoid the presence of states with negative frequencies. The frequencies of the relevant scattering modes can thus be labelled with integer indices: denoting by $\lceil x \rceil$ the lowest integer larger than $x$, these frequencies are $\omega_q = \omega_0 - \Omega_0 \times \lceil \omega_0/\Omega_0 \rceil + q\Omega_0$ with $q > 0$.

In one-dimensional geometry, two different travelling-wave modes of the form $\exp[i(kz - \omega t)]$ can have the same frequency $\omega = c|k|$, corresponding to positive $k$ (right-moving) and to negative $k$ (left-moving). However, phase modulators are ideally reflectionless devices, so that right-moving incoming radiation does not couple to left-moving outgoing radiation on the same side of the modulator. The scattering is thus fully described by an operator transforming right-moving modes into right-moving modes, together with an operator acting only on left-moving modes. These two operators are different in general, since the properties of a phase modulator in reverse operation depends on the electrodes' configuration. However, both operators should be of the same form, and so we will restrict to right-moving modes. Then, we can unambiguously label these relevant modes by its frequency $\omega$ or frequency index $q > 0$. For instance, a one-photon state at such a right-travelling mode can be equivalently described as $|1_\omega\rangle$ or as $|1_q\rangle$. In addition, conventional electro-optic modulators are also designed for a unique input polarization, and therefore it is not necessary to add a new index for polarization.

The multimode Hilbert space of right-moving travelling wave modes with frequencies $\omega_q$ ($q = 1, 2,..$) will be denoted by $\mathcal{F}_+$, and it is the space upon which the phase modulator will be defined. This space can be decomposed in subspaces of definite number of photons as:

$$\mathcal{F}_+ = \bigoplus_{N=0}^{\infty} \mathcal{F}_N \tag{3.11}$$

where $\mathcal{F}_N$ is the subspace of Fock states having $N$ photons, which is spanned by vectors:

$$\left| (n_1)_{q_1} (n_2)_{q_2} \cdots (n_M)_{q_M} \right\rangle = \prod_{s=1}^{M} \frac{(\hat{a}_{q_s}^+)^{n_s}}{\sqrt{n_s!}} |vac\rangle \qquad \sum_{s=1}^{M} n_s = N \qquad (3.12)$$

where $q_s$ is the mode index, $M$ is the number of occupied modes and $n_s$ the corresponding occupation number. For instance, $\mathcal{F}_0$ is composed by the vacuum sate, $|vac\rangle$, whereas the subspace $\mathcal{F}_1$ is generated by all one-photon states $|1_q\rangle = \hat{a}_q^+ |vac\rangle$ at frequency $\omega_q$ ($q>0$).

The scattering operator describing single-tone phase modulation, analogous to Eq. (3.7), has been constructed and solved in [63,64]. It is a unitary multimode operator of the form $\hat{S}_{PM} : \mathcal{F}_+ \to \mathcal{F}_+$ given by:

$$\hat{S}_{PM} = \exp\left(i\varphi_B \hat{N} - i\tfrac{1}{2} m e^{-i\theta} \hat{T} - i\tfrac{1}{2} m e^{i\theta} \hat{T}^+\right) = \exp(i\varphi_B \hat{N}) \exp(-i\hat{G}) \qquad (3.13)$$

where the $\hat{T}$ operator raises the mode of photons, $\hat{N}$ is the photon number:

$$\hat{T} = \sum_{n=1}^{\infty} \hat{a}_{n+1}^+ \hat{a}_n \qquad\qquad \hat{N} = \sum_{n=1}^{\infty} \hat{a}_n^+ \hat{a}_n \qquad (3.14)$$

and having used that $[\hat{N},\hat{T}] = 0$ to decompose the operator and introduce the effective Hamiltonian $\hat{G}$. Equation (3.13) can be justified either form the comparison with the perturbative generation of the Bessel function describing the classical transition amplitudes [63] or from physical considerations. To this end, first observe that the positive-frequency part of the cosine modulation in Eq. (3.6) simply increases the mode frequency by an amount $\Omega_0$, whereas the negative-frequency part decreases it. As in the classical picture, the strength of this mode coupling is proportional to the amplitude $V_m$ of the radio-frequency mode propagating along the modulator's electrodes, which is assumed classical (not quantized). Equivalently, the electro-optic interaction can be described by a three-boson coupling between two optical modes and a RF mode in a coherent state which is set to its average value $V_m$ [70]. This results in a coupling constant that equals the classical modulation index $m$, and can be justified from the high value of the electric field in the electro-optic material, which is of the order of kVm$^{-1}$ in bulk-optics modulators and even higher in integrated modulators. Alternative approaches to the electro-optic coupling include a quantized microwave field together with a finite number of optical modes [71-73].

As for the bias term in Eq. (3.13), it assumes absence of dispersion, so that all modes undergo the same propagation phase. Dispersive propagation can be introduced simply by changing the first term in the exponential by $\hat{\varphi}_B = \Sigma_n \varphi_{Bn} \hat{a}_n^+ \hat{a}_n$. This operator no longer commutes with $\hat{G}$, so the separation in the rightmost part of Eq. (3.13) does not hold in general. Due to the small physical length of the modulators, effects due to dispersive propagation of optical modes inside the device are usually neglected. Indeed, it is easy to show that $[\hat{\varphi}_B, \hat{G}] = 0$ if dispersion can be neglected at the scales of $\Omega_0$, i.e., $\varphi_{Bn} \cong \varphi_{B,n\pm1}$. The main consequence of dispersion inside the modulator is the decrease of modulation bandwidth due to walk-off between the microwave in the modulator's electrode and optical modes in the waveguide, an effect that will be analyzed in the following subsection. In any event, even if dispersion is included, the interaction preserves the number of incoming photons, $[\hat{N}, \hat{S}_{PM}] = 0$.

Remarkably, the theory can be exactly solved neglecting dispersive propagation. Clearly, operator $\hat{S}_{PM}$ leaves the vacuum invariant, $\hat{S}_{PM}|vac\rangle = |vac\rangle$, so that the action of the scattering operator in the Schrödinger picture over an arbitrary state in the Fock basis is reduced to the computation of its adjoint action over an arbitrary photon creation operator: $\hat{S}_{PM} \hat{a}_p^+ ... \hat{a}_k^+ |vac\rangle = \hat{S}_{PM} \hat{a}_p^+ \hat{S}_{PM}^+ ... \hat{S}_{PM} \hat{a}_k^+ \hat{S}_{PM}^+ |vac\rangle$. This computation has been carried out in [63] using a diagrammatic technique and in [64] using the spectral decomposition of $\hat{S}_{PM}$, which can be derived by analogy with the well-known spectrum of the single-mode Susskind-Glogower cosine operator. The result is:

$$\hat{S}_{PM} \hat{a}_k^+ \hat{S}_{PM}^+ = e^{i\varphi_B} \sum_{q=1}^{\infty} e^{-i\theta(q-k)} \left[ (-i)^{q-k} J_{q-k}(m) - (-i)^{q+k} J_{q+k}(m) \right] \hat{a}_q^+ \qquad (3.15)$$

and its action on one-photon states is therefore:

$$\hat{S}_{PM} |1_q\rangle = e^{i\varphi_B} \sum_{k=1}^{\infty} e^{-i\theta(k-q)} \left[ (-i)^{k-q} J_{k-q}(m) - (-i)^{k+q} J_{k+q}(m) \right] |1_k\rangle \equiv \sum_{k=1}^{\infty} S_{k,q} |1_k\rangle \qquad (3.16)$$

where the one-photon scattering coefficients from mode $q$ to mode $k$, $S_{k,q} = \langle 1_k | \hat{S} | 1_q \rangle$ have been introduced, which form an unitary matrix $(S^{-1})_{k,q} = S_{q,k}^*$. These matrix elements have the interpretation of probability amplitudes for a process leading to the creation of a photon in mode $k$ from a photon at mode $q$. The central role played by $S_{k,q}$ can be appreciated by noticing that an arbitrary scattering amplitude is entirely determined by these quantities. This is shown by the normal-order form of the scattering operator:

$$\hat{S}_{PM} =: \exp[\sum_{k,q=1}^{\infty} \hat{a}_k^+ (S_{k,q} - \delta_{k,q}) \hat{a}_q ]: \tag{3.17}$$

The proof of this formula is standard [74]: first, it is calculated the action over an arbitrary multimode displacement operator, which gives:

$$\hat{S}_{PM} \bigotimes_{q=1}^{\infty} \hat{D}_q(\alpha_q) \hat{S}_{PM}^+ = \bigotimes_{k=1}^{\infty} \hat{D}_k (\sum_{q=1}^{\infty} S_{k,q} \alpha_q) \tag{3.18}$$

From this, the Husimi function is readily computed, and Eq. (3.17) follows from the optical equivalence theorem [59]. Equation (3.18) also formalizes the correspondence between one-photon scattering amplitudes and coherent-sate scattering amplitudes.

Now, the classical expression in Eq. (3.8) can be recovered using an appropriate approximation. First the operator is decomposed as $\hat{S}_{PM} = \hat{S}_{PM}^{(1)} - \hat{S}_{PM}^{(2)}$ [64], with matrix elements:

$$S_{k,q}^{(1)} = e^{i\varphi_B} e^{-i\theta(k-q)} (-i)^{k-q} J_{k-q}(m) \qquad S_{k,q}^{(2)} = e^{i\varphi_B} e^{-i\theta(k-q)} (-i)^{k+q} J_{k+q}(m) \tag{3.19}$$

The first contribution $S_{k,q}^{(1)}$ provides the leading amplitudes in the optical limit, i.e, when input modes belong to optical bands ($q \gg 1$), and coincide with the transition amplitudes to sideband in the classical formalism, as given by Eq. (3.8): $S_{k,q}^{(1)} = C_{k-q}$. The second contribution $S_{k,q}^{(2)}$ is a correction required by unitarity and the fact that modes have positive frequencies ($k, q > 0$). More formally, the optical limit amounts to an approximation of the form:

$$\hat{S}_{PM} |\Psi\rangle \cong \hat{S}_{PM}^{(1)} |\Psi\rangle \tag{3.20}$$

for states $|\Psi\rangle$ whose spectral content is contained in sufficiently large mode indices $q$. For instance, for one-photon states $|1_q\rangle$ the norm of the error term in Eq. (3.20) tends to zero as $q \to \infty$:

$$\| \hat{S}_{PM}^{(2)} |1_q\rangle \|^2 = \sum_{k=1}^{\infty} |J_{k+q}(m)|^2 \leq \sum_{k=1}^{\infty} \frac{1}{(k+q)!^2} \left(\frac{m}{2}\right)^{2k+2q}$$
$$\leq \frac{1}{(q+1)!^2} \left(\frac{m}{2}\right)^{2q+2} \sum_{n=0}^{\infty} \frac{1}{n!^2} \left(\frac{m}{2}\right)^{2n} = \frac{I_0(m)}{(q+1)!^2} \left(\frac{m}{2}\right)^{2q+2} \xrightarrow[q \to \infty]{} 0 , \tag{3.21}$$

where $I_0$ is the modified Bessel function of the first kind and order zero and where the bound $|J_s(m)| \leq (m/2)^s/s!$ [75] and that $(a+b)! \geq a!b!$ have been used. Then, the optical limit amounts to any of the following expressions [25, 29]:

$$\hat{S}_{PM} |1_q\rangle \cong \hat{S}_{PM}^{(1)} |1_q\rangle \cong e^{i\varphi_B} \sum_{s=-\infty}^{\infty} (-ie^{-i\theta})^s J_s(m) |1_{q+s}\rangle = e^{i\varphi_B} \sum_{s=-\lceil m \rceil-1}^{\lceil m \rceil+1} (-ie^{-i\theta})^s J_s(m) |1_{q+s}\rangle \tag{3.22}$$

The first expression is the direct translation of the classical coefficients in Eq. (3.6), where the lower $s$ limit is extended to $-\infty$ since the corresponding amplitudes $J_s(m)$ are negligible. The second expression in Eq. (3.22) follows from Carson's rule [69], which amounts to the observation that the Bessel functions $J_s(m)$ can be neglected for indices $s > \lceil m \rceil + 1$. This last expression explicitly shows that the number of relevant sidebands generated by phase modulation from a single carrier is of the order of $m$.

*3.3 Arbitrary modulation in continuous-mode formalism*

The extension of the previous theory to arbitrary modulation inevitably requires the introduction of a continuous-mode formalism. In this case, the modulating voltage is parametrized as $V(t) = V_{DC} + \Delta V(t)$ with $\Delta V(t) = V_m x(t)$. Here, function $x(t)$ is the (dimensionless) modulation function, which is assumed real, dc-free (the dc-term is explicitly separated in $V_{DC}$), dc-balanced (that is, $\int x(t) dt = 0$) and bandlimited. The modulation index in then described in terms of the level $V_m$ as before, $m = \pi V_m/V_\pi$. Under these conditions, $x(t)$ can be expressed in terms of its Fourier transform as:

$$x(t) = \int_{-\infty}^{+\infty} \frac{d\Omega}{2\pi} X(\Omega) e^{-i\Omega t} \qquad X(-\Omega) = X(\Omega)^* \tag{3.23}$$

with $X(-\Omega) = X(\Omega)^*$ and $X(0) = 0$. Then, the effective Hamiltonian associated to this function is:

$$\hat{G} = m \int_0^{+\infty} \frac{d\omega}{2\pi} \int_0^{+\infty} \frac{d\Omega}{2\pi} \left[ X(\Omega)\hat{a}^+(\omega+\Omega)\hat{a}(\omega) + X(\Omega)^* \hat{a}^+(\omega)\hat{a}(\omega+\Omega) \right] \quad (3.24)$$

which is manifestly hermitian and involves positive frequencies only. In Eq. (3.24) the continuous-mode creation and annihilation operators have been introduced with the following commutation relation:

$$[\hat{a}(\omega), \hat{a}^+(\omega')] = 2\pi\delta(\omega-\omega') \quad (3.25)$$

The normalization with a $2\pi$ factor has been chosen to conform to the Fourier transform convention of Eq. (3.23). Effects of modulation bandwidth can be easily included in Eq. (3.24) by use of the substitution $X(\Omega) \to H(\Omega)X(\Omega)$, where $H(\Omega)$ is the radio-frequency transfer function of the modulator. In what follows it is assumed that $H(\Omega)=1$ for ease of notation. Then, the phase modulation unitary operator is

$$\hat{S}_{PM} = \exp(i\hat{\varphi}_B - i\hat{G}) \quad (3.26)$$

with operator

$$\hat{\varphi}_B = \int_0^{+\infty} \frac{d\omega}{2\pi} \varphi_B(\omega) \hat{a}^+(\omega)\hat{a}(\omega) \quad (3.27)$$

describing the (eventually dispersive) propagation phase inside the modulator. For instance, single-tone modulation in continuous-mode formalism is described by a driving field of the form:

$$X(\Omega) = \frac{1}{2} e^{-i\theta}\delta(\Omega-\Omega_0) + \frac{1}{2} e^{i\theta}\delta(\Omega+\Omega_0) \quad \Omega_0 > 0 \quad (3.28)$$

so that

$$\hat{G} = \frac{m}{2} \int_0^{+\infty} \frac{d\omega}{2\pi} e^{-i\theta} \hat{a}^+(\omega+\Omega_0)\hat{a}(\omega) + hc \quad (3.29)$$

which is the continuous-mode generalization of Eq. (3.13)

The action of operator $\hat{S}_{PM}$ in Eq. (3.26) cannot be solved for arbitrary modulating function and, as in the classical theory of phase modulation [69], it is necessary some kind of approximation such as a perturbative expansion in modulation index $m$. This expansion can be presented as:

$$\hat{S}_{PM} \cong e^{i\hat{\varphi}_B} e^{-i\hat{G}} \cong e^{i\hat{\varphi}_B}(1 - i\hat{G}) \quad (3.30)$$

Here the exponential has been expanded to first order and it is assumed that dispersion can be neglected at the scale of the modulation bandwidth, so $[\hat{\varphi}_B, \hat{G}]=0$. For instance, under multi-tone modulation:

$$X(\Omega) = \sum_k \frac{A_k e^{-i\theta_k}}{2} \delta(\Omega-\Omega_k) + \sum_k \frac{A_k e^{i\theta_k}}{2} \delta(\Omega+\Omega_k) \quad \Omega_k > 0 \quad (3.31)$$

so that

$$\hat{S}_{PM} \cong e^{i\hat{\varphi}_B} \left[ \hat{1} - im \sum_k \frac{A_k e^{-i\theta_k}}{2} \int_0^{+\infty} \frac{d\omega}{2\pi} \hat{a}^+(\omega+\Omega_k)\hat{a}(\omega) - hc \right] \quad (3.32)$$

A different development of operator in Eq. (3.26) follows from an approximate summation of the perturbative series that explicitly neglects effects due to the existence of a lower bound in frequencies. This development can be formulated as the quantum equivalent to the classical expression in Eq. (3.5) of the action of phase modulation over the complex envelope, so the complex envelope operator [47] is first introduced. In continuous-mode formalism, the positive-frequency part of the electric-field operator of dispersive media characterized by a refractive index $\eta(\omega)$ is given, in Heisenberg image, by [74]:

$$\hat{E}^{(+)}(z,t) = i \int_0^{\infty} \frac{d\omega}{2\pi} \left( \frac{\hbar\omega}{2\varepsilon_0 c\eta(\omega)S} \right)^{1/2} \hat{a}(\omega) \exp[ik(\omega)z - i\omega t] \quad (3.33)$$

with $k(\omega) = \omega\eta(\omega)/c$ and $S$ is the quantization area. The slowly varying envelope operator is defined by

$$\hat{A}(z,t) = \int_0^\infty \frac{d\omega}{2\pi} \hat{a}(\omega) \exp\left[i(k(\omega) - k(\omega_0))z - i(\omega - \omega_0)t\right] \qquad (3.34)$$

and represents a description of the electric field for radiation confined to a band around a central frequency $\omega_0$. Indeed, under this condition one can approximate $\omega \approx \omega_0$ in the square root of the integrand of Eq. (3.33), leading to:

$$\hat{E}^{(+)}(z,t) \propto \hat{A}(z,t) \exp[ik(\omega_0)z - i\omega_0 t] \qquad (3.35)$$

which is the quantum counterpart of Eq. (3.4). The action of the phase modulator over operator $\hat{A}(z,t)$ will be assumed to occur at $z = 0$ so that the relevant computation, in the Heisenberg image, is:

$$\begin{aligned}
\hat{S}_{PM}^+ \hat{A}(0,t)\hat{S}_{PM} &= \exp(i\omega_0 t)\int_0^\infty \frac{d\omega}{2\pi} \hat{S}_{PM}^+ \hat{a}(\omega) \hat{S}_{PM} \exp(-i\omega t) \\
&= \exp(i\omega_0 t + i\varphi_B)\int_0^\infty \frac{d\omega}{2\pi} e^{i\hat{G}} \hat{a}(\omega) e^{-i\hat{G}} \exp(-i\omega t)
\end{aligned} \qquad (3.36)$$

where the dispersion inside the modulator for optical inputs at the scales of the modulation bandwidth has been neglected so that $\hat{S}_{PM} = \exp(i\hat{\varphi}_B)\exp(-i\hat{G})$. Now, the exponential series of $\exp(i\hat{G})\hat{a}(\omega)\exp(-i\hat{G})$ involves commutators of the form:

$$[i\hat{G},...[i\hat{G}, \hat{a}(\omega)]...]^{k)} = (-im)^k \int_0^{+\infty} \frac{d\Omega_0}{2\pi} \int_0^{+\infty} \frac{d\Omega_1}{2\pi} ... \int_0^{+\infty} \frac{d\Omega_k}{2\pi} \times$$
$$\left[\sum_{n=0}^k \binom{k}{n} X(\Omega_0)X(\Omega_1)...X(\Omega_n)X(\Omega_{n+1})^* ... X(\Omega_k)^* \hat{a}(\omega - \Omega_0 - \Omega_1 - ... - \Omega_n + \Omega_{n+1} + ... + \Omega_k\right] \qquad (3.37)$$

This expression is, however, incorrect, as it allows negative-frequency modes in the argument of the creation operator. This issue can be neglected by assuming that the error introduced by formally considering negative-frequency modes is small when applied to optical fields whose spectrum is contained in the vicinity of the optical frequency $\omega_0$ or, in other words, when the radiation is narrowband around $\omega_0$ [57, 74]. The easiest way of introducing this approximation is to extend the lower integration limit in Eq. (3.34) up to $-\infty$ [57, 74]. Then, inserting this perturbative expansion in Eq. (3.36) and integrating over $\omega$, yields a term proportional to operator $\hat{A}(0,t)$ that can be factored out. The remaining $k+1$ integrals in $\Omega_n$ ($n = 0...k$) at each perturbative level $k$ can be performed after noticing that

$$\int_0^{+\infty} \frac{d\Omega}{2\pi} X(\Omega)e^{-i\Omega t} = \frac{1}{2}\tilde{x}(t) \qquad (3.38)$$

where $\tilde{x}(t)$ is the analytic signal associated to $x(t)$. The perturbative series is then:

$$\begin{aligned}
\hat{S}_{PM}^+ \hat{A}(0,t)\hat{S}_{PM} &\cong \hat{A}(0,t)\exp(i\varphi_B)\sum_{k=0}^\infty \frac{1}{k!}\left(-i\frac{m}{2}\right)^k \sum_{n=0}^k \binom{k}{n} \tilde{x}(t)^n [\tilde{x}(t)^*]^{k-n} \\
&= \hat{A}(0,t)\exp(i\varphi_B)\sum_{k=0}^\infty \frac{1}{k!}\left(-i\frac{m}{2}\right)^k [\tilde{x}(t) + \tilde{x}(t)^*]^k = \hat{A}(0,t)\exp[i\varphi_B - imx(t)]
\end{aligned} \qquad (3.39)$$

where $x(t) = \operatorname{Re}\tilde{x}(t)$ has been used. The final expression is the quantum analogue of Eq. (3.5).

An alternative yet equivalent formulation of Eq. (3.39) is based on the Fourier-transformed continuous-mode operator [57, 74],

$$\hat{a}(t) = \int_{-\infty}^\infty \frac{d\omega}{2\pi} e^{-i\omega t} \hat{a}(\omega) = \hat{A}(t)e^{-i\omega_0 t} \qquad (3.40)$$

with commutation relations:

$$[\hat{a}(t), \hat{a}^+(t')] = \delta(t - t') \qquad (3.41)$$

The formal extension to negative-frequency modes in Eq. (3.40) represents, as above, a negligible error only when this operator acts over states whose spectral content lies in optical bands. Comparison with Eq. (3.39) shows that operator $\hat{a}(t)$ transforms in the same manner as $\hat{A}(t)$ upon phase modulation.

## 4. Quantum model of the Electro-optic Amplitude Modulator

*4.1 Introductory remarks*

The electro-optic amplitude modulator (EOM) is the most popular modulating device employed in high speed optical communications systems featuring line rates in excess of 2.5 Gb/s [1, 76]. It is also widely employed in analog photonic applications [2] and radio over fiber systems [10, 12] where the modulating subcarriers are located in the RF, microwave and millimeter–wave regions of the electromagnetic spectrum. The integrated EOM is built by embedding one or two electro-optic phase modulators into a Mach-Zehnder interferometric setup closed by two guided-wave beamsplitters`[12, 4] as shown in figure. 4.1. There are however various forms to assemble the interferometer. For instance, in figure 4.2 the most common EOM designs that are encountered in practice are shown [12]. EOMs with only one internal phase modulator such as B), D), F) in figure 4.2 are known as single drive or asymmetric modulators, while EOMs with two internal phase modulators, such as A), C) and E) in figure 4.2 are known as dual drive modulators. For each phase modulator there are two ports, one for the modulator dc bias voltage and another to inject the time varying modulating signal. Designs A) and B) correspond to Y-Branch modulators, where both the input and output beamsplitters in the Mach-Zehnder interferometric structure are Y-Branch power splitters. These are the most common commercial devices. Designs C) and D) represent the so-called DC (directional coupler) modulators where both the input and output beamsplitters in the Mach-Zehnder interferometric structure are guided wave directional couplers. These modulators bring the added value of an extra input and an extra output port, but are more expensive to produce since the fabrication of a symmetric directional coupler is more challenging as compared to the Y-Branch power splitter which is particularly simple to implement in integrated fashion [66, 67]. DC modulators were common in the early days of integrated optics but today, although available upon request from different vendors they are not a mainstream commercial product. Finally, designs E) and F) correspond to hybrid Y-Branch/DC modulators which feature one input and two outputs. These modulators, which can be found upon request in the market, are common for Cable TV applications [12], where the dual output permits a first signal splitting in the broadcasting header. The operation and design principles of the EOM under classical conditions are quite well established and understood and the interested reader can find useful information in innumerable references in the literature [2, 4, 5, 68]. Its quantum operation has been recently analyzed in [65] and the salient features are described in the remaining of this section. Of course, as in the case of phase modulators, bulk optics configurations are possible, in this case by inserting a bulk optics phase modulator in between two bulk polarizers as shown in [4, 5, 68].

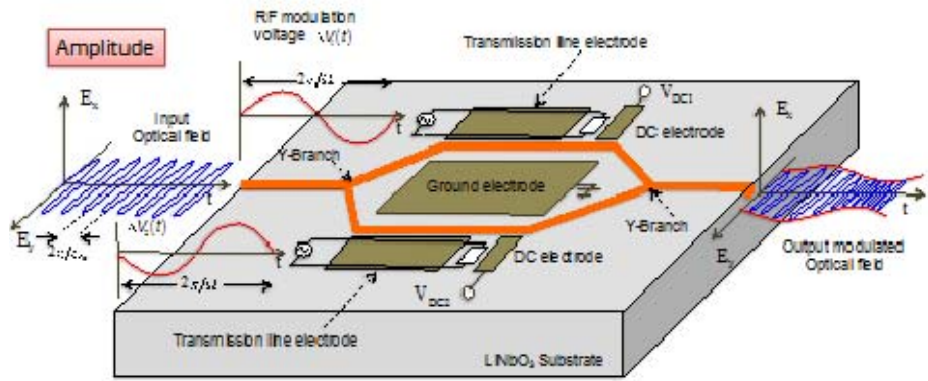

**Figure 4.1.** Layout of a dual-drive, Y-Branch electro-optic amplitude modulator.

*4.2 General Model for Electro-optic Amplitude Modulators*

Let us consider the general structure of the amplitude modulator as shown in figure 4.3. The layout is obviously an abstract representation as the beamsplitters are of the guided wave type as discussed in section 2. However, the representation of figure 4.3 clearly shows the three building blocks of the modulator; an input beamsplitter opening the two paths of the interferometer, two different paths (labeled 1 and 2 in the figure) which include each one a phase modulator, and an output beamsplitter which closes the interferometer and provides two possible output ports. Its classical description is readily extracted from figure 4.3. If input and output (with primes) waves are described by their complex envelopes, the input/output relationship can be written in matrix form as.

$$\begin{pmatrix} A_1'(t) \\ A_2'(t) \end{pmatrix} = \begin{pmatrix} m_{11}(t) & m_{12}(t) \\ m_{21}(t) & m_{22}(t) \end{pmatrix} \cdot \begin{pmatrix} A_1(t) \\ A_2(t) \end{pmatrix} \tag{4.1}$$

where $m_{ab}(t)$ represents the classical modulation function of a wave entering in port $b$ and exiting by port $a$, which are given by:

$$\begin{pmatrix} m_{11}(t) & m_{12}(t) \\ m_{21}(t) & m_{22}(t) \end{pmatrix} = \begin{pmatrix} t_o' & r_o \\ r_o' & t_o \end{pmatrix} \cdot \begin{pmatrix} e_1(t) & 0 \\ 0 & e_2(t) \end{pmatrix} \cdot \begin{pmatrix} t_i' & r_i \\ r_i' & t_i \end{pmatrix} = \begin{pmatrix} t_o'e_1t_i' + r_oe_2r_i' & t_o'e_1r_i + r_oe_2t_i \\ r_o'e_1t_i' + t_oe_2r_i' & r_o'e_1r_i + t_oe_2t_i \end{pmatrix} \tag{4.2}$$

with $e_k(t)=\exp[i\varphi_{Bk}-imx_k(t)]$, $k=1,2$, the phase impinged by each phase modulator. This matrix is unitary, which implies that:

$$\begin{aligned}|m_{11}(t)|^2 + |m_{21}(t)|^2 &= |m_{22}(t)|^2 + |m_{12}(t)|^2 = 1 \\ m_{11}(t)m_{12}(t)^* + m_{21}(t)m_{22}(t)^* &= m_{11}(t)m_{21}(t)^* + m_{12}(t)m_{22}(t)^* = 0 \end{aligned} \tag{4.3}$$

These classical expressions will be of use further below.

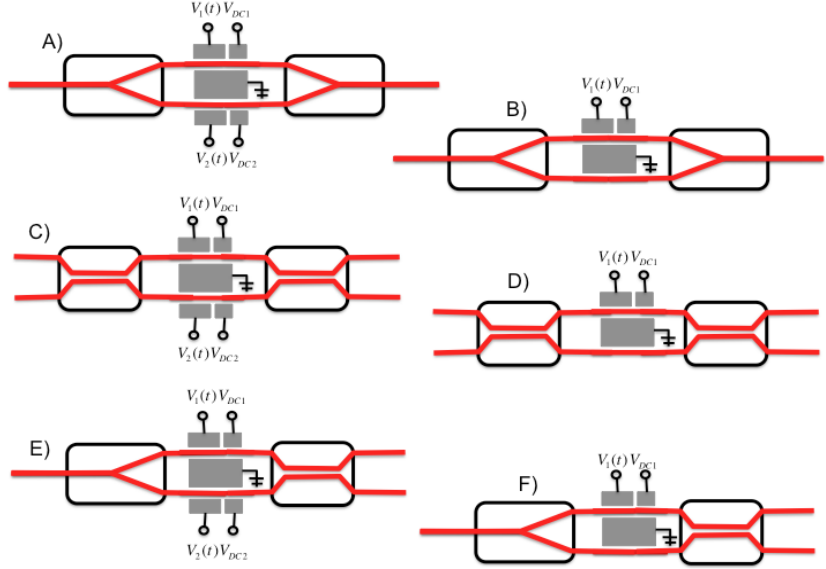

**Figure 4.2.** Different possible layouts for amplitude electro-optic modulators. After [12]

As for the quantum device, the operator describing the action of the EOM is given by:

$$\hat{S}_{EOM} = \hat{S}_{BS_o}\hat{S}_{PM}\hat{S}_{BS_i} = \hat{S}_{BS_o}(\hat{S}_{PM_1} \otimes \hat{S}_{PM_2})\hat{S}_{BS_i} \tag{4.4}$$

Here $\hat{S}_{BSi}$ and $\hat{S}_{BSo}$ represent the scattering operators of the input and output beamsplitters respectively and $\hat{S}_{PM\,1}, \hat{S}_{PM\,2}$ represent the scattering operators corresponding, respectively, to the phase modulators located in the lower and upper paths of the layout of figure 4.3 and which have been discussed thoroughly in section 3. $\hat{S}_{EQM}$ is unitary, since it is the product of three unitary operators, so that $\hat{S}_{EOM}\hat{S}^+_{EOM} = \hat{S}^+_{EOM}\hat{S}_{EOM} = \hat{1} \otimes \hat{1}$. It should be also pointed out that although this complete layout is used to construct the model for the amplitude modulator, the equations developed here can and will be particularized to the case of asymmetric modulators as well where, say, $\hat{S}_{PM\,1}$ is substituted by the identity operator. Equation (4.4) is completely general and can be employed to analyze the operation of the electro-optic amplitude modulator under different working conditions. Different approximations can be introduced through the form of the phase modulation operators.

An important representative case is that corresponding to the characterization of the EOM response to single photon inputs. For instance, for a photon at mode $n$ in the first port:

$$\hat{S}_{EOM}\left(|1\rangle_n \otimes |vac\rangle\right) = \hat{S}_{EOM}\left(\hat{a}_n^+ \otimes \hat{1}\right)|vac\rangle \otimes |vac\rangle = \hat{S}_{EOM}\left(\hat{a}_n^+ \otimes \hat{1}\right)\hat{S}^+_{EOM}|vac\rangle \otimes |vac\rangle \tag{4.5}$$

and use of the results derived in section 2 and (4.4) yields:

$$\hat{S}_{EOM}\left(\hat{a}_n^+ \otimes \hat{1}\right)\hat{S}^+_{EOM} = t_i't_o'\left(\hat{b}_n^+ \otimes \hat{1}\right) + r_i'r_o\left(\hat{c}_n^+ \otimes \hat{1}\right) + t_i'r_o'\left(\hat{1} \otimes \hat{b}_n^+\right) + r_i't_o\left(\hat{1} \otimes \hat{c}_n^+\right) \tag{4.6}$$

where $\hat{b}_n^+ \equiv \hat{S}_{PM_1}\hat{a}_n^+\hat{S}_{PM_1}^+$ and $\hat{c}_n^+ \equiv \hat{S}_{PM_2}\hat{a}_n^+\hat{S}_{PM_2}^+$ have the interpretation of photon creators by phase modulation. Equation (4.6) defines a transformation with a straightforward physical interpretation, analogous to the classical case. For instance, the first term in the right member represents the transformation corresponding to a state entering the EOM through port 1, transmitted by the input beamsplitter $BS_i$ to the lower path in figure 4.3, being modulated by $PM_1$ and finally transmitted by the output beamsplitter $BS_o$ to output port 1. Similarly, for a photon in the second port:

$$\hat{S}_{EOM}^+\left(\hat{1}\otimes\hat{a}_n^+\right)\hat{S}_{EOM} = r_i't_o'\left(\hat{b}_n^+\otimes\hat{1}\right) + t_ir_o\left(\hat{c}_n^+\otimes\hat{1}\right) + r_i'r_o'\left(\hat{1}\otimes\hat{b}_n^+\right) + t_it_o\left(\hat{1}\otimes\hat{c}_n^+\right) \quad (4.7)$$

with a similar physical interpretation for its four terms but taking into account that now the input is in port 2. Although Eqs. (4.6) and (4.7) have been derived for a perfectly balanced interferometer they hold for unbalanced structures since any phase imbalance between the upper and the lower arms can be incorporated into the dc bias terms $\varphi_B$.

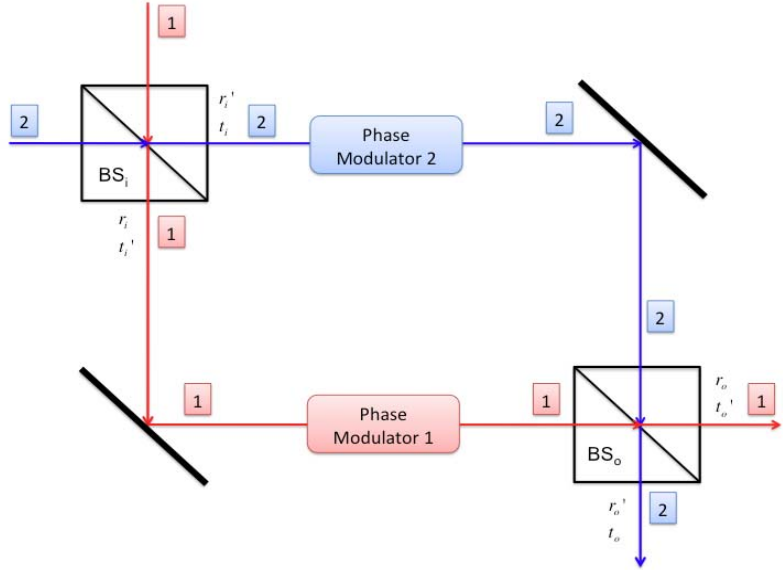

**Figure 4.3.** Generic layout of an electro-optic amplitude modulator

As with phase modulation, the above expressions can be extended to continuous-mode formalism. A complete theory can be developed in parallel to the analysis in the previous section, starting from the definition of a complex envelope operator for each of the two input/output ports and then transforming them to the Heisenberg image as with the phase modulator. For instance, the transformed first-port complex envelope operator at $z = 0$ is:

$$\hat{S}_{EOM}^+\left(\hat{A}(0,t)\otimes\hat{1}\right)\hat{S}_{EOM} = \exp(i\omega_0 t)\int_0^\infty \frac{d\omega}{2\pi}\hat{S}_{EOM}^+\left(\hat{a}(\omega)\otimes\hat{1}\right)\hat{S}_{EOM}\exp(-i\omega t) \quad (4.8)$$

This transformation can be computed using Eq. (4.4), the action of beamsplitters, Eq. (2.1), and the action of phase modulators in continuous-mode formalism, Eq. (3.36), under the approximation valid for narrowband optical states. The computation yields:

$$\hat{S}_{EOM}^+\left(\hat{A}(0,t)\otimes\hat{1}\right)\hat{S}_{EOM} = m_{11}(t)\left(\hat{A}(0,t)\otimes\hat{1}\right) + m_{12}(t)\left(\hat{1}\otimes\hat{A}(0,t)\right)$$
$$\hat{S}_{EOM}^+\left(\hat{1}\otimes\hat{A}(0,t)\right)\hat{S}_{EOM} = m_{21}(t)\left(\hat{A}(0,t)\otimes\hat{1}\right) + m_{22}(t)\left(\hat{1}\otimes\hat{A}(0,t)\right) \quad (4.9)$$

Then, as in phase modulation, the transformation of the complex envelopes in the Heisenberg image is analogous to its classical description in Eq. (4.2).

## 5. Selected Applications

### 5.1 Introductory remarks

The results and models developed in sections 3 and 4 for phase and amplitude modulation are general in the sense that they ultimately depend on the particular formulation that is chosen for the phase modulation operation. These formulations can be classified in order of increasing refinement. First, Eq. (3.15) is a unitary discrete-mode representation of PM under single tone modulation. Its approximation described by Eq. (3.22) allows for the presence of negative frequencies, and therefore is valid only for input radiation in optical bands. And secondly, Eq. (3.26) is a continuous-mode representation of PM under arbitrary driving voltage. This model is only tractable in perturbative series, Eq. (3.30), or, permitting again negative frequencies, as an effective model for the complex envelope operator when acting over narrowband optical radiation, as is described by Eq. (3.39). Also, the modulating field can also be categorized into single discrete mode (sinusoidal RF) modulation, multiple discrete mode (subcarrier) modulation and continuous-mode or pulsed modulation. This section analyzes specific examples of some of the most representative applications of both phase and amplitude electro-optic modulators.

*5.2 Y-Branch Electro-optic Amplitude Modulator with two sinusoidal modulating inputs: Double and Single Sideband Modulation of single photon states.*

The first application shows the use of amplitude modulators as voltage-controlled scattering devices for providing user-defined transition probabilities to the sidebands of one-photon states through single-tone modulation. The layout of this modulator is shown in figure 4.2A). This configuration is widely employed in multiple applications within the field of RF Photonics, and in the quantum context it has been used in QKD systems for sideband generation. In practical devices the Y-Branch design is such that:

$$t_o = t_o^{'} = t_i = t_i^{'} = 1/\sqrt{2} \qquad r_i^{'} = r_o = -r_i = -r_o^{'} = 1/\sqrt{2} \qquad (5.1)$$

It will be also assumed that the microwave frequency $\Omega_0$ of the sinusoidal modulating signals in both arms of the interferometer is the same, so that the control of the transition amplitudes is obtained by bias $\varphi_B$ and radio-frequency delay $\theta$. The relevant model for this analysis is the discrete-mode model in Eq. (3.16), from which it can be employed its approximation in Eq. (3.22) if convenient. Since $\Omega_0$ is the same for both PM modulators the frequencies of the sideband modes generated are the same, and thus the action over a given mode $n$ can be compactly written as:

$$\hat{b}_n^+ \equiv \hat{S}_{PM_1} \hat{a}_n^+ \hat{S}_{PM_1}^+ = \sum_{q=1}^{\infty} S_{q,n} \hat{a}_q^+ \qquad \hat{c}_n^+ \equiv \hat{S}_{PM_2} \hat{a}_n^+ \hat{S}_{PM_2}^+ = \sum_{q=1}^{\infty} \overline{S}_{q,n} \hat{a}_q^+ \qquad (5.2)$$

where the overbar in the second expressions means that the parameters defining this phase modulator can be different from the first. Then, the response to a single photon input is given by:

$$\hat{S}_{EOM} |1\rangle_n \otimes |vac\rangle = \frac{1}{2}\left[\sum_{q=1}^{\infty}(S_{q,n} + \overline{S}_{q,n})|1\rangle_q \otimes |vac\rangle\right] + \frac{1}{2}\left[|vac\rangle \otimes \sum_{q=1}^{\infty}(-S_{q,n} + \overline{S}_{q,n})|1\rangle_q\right] \qquad (5.3)$$

Now, this output can be specialized for two standard settings of this modulator through the definition of bias, modulation indices and radio-frequency phase. The fist case is the so called *Double Sideband (DSB) operation in quadrature* [10,12] which, due to the low harmonic distortion represents one of the most standard settings in analog communications. It is characterized by $m_1 = m_2 = m$, $\theta_1 = 0$, $\theta_2 = \pi$, $\varphi_{B1} = -\varphi_{B2} = \pi/2$. Using Eq. (3.15):

$$S_{q,n} = i\left[(-i)^{q-n} J_{q-n}(m) - (-i)^{q+n} J_{q+n}(m)\right]$$
$$\overline{S}_{q,n} = -i\left[i^{q-n} J_{q-n}(m) - i^{q+n} J_{q+n}(m)\right] = S_{q,n}^* \qquad (5.4)$$

so Eq. (5.3) becomes, after use of Eq. (5.2):

$$\hat{S}_{EOM}\left(\hat{a}_n^{\dagger} \otimes \hat{1}\right)\hat{S}_{EOM}^+ = \sum_{q=1}^{\infty} \text{Re}(S_{q,n}) \hat{a}_q^+ \otimes \hat{1} + \hat{1} \otimes \sum_{q=1}^{\infty} -i\,\text{Im}(S_{q,n}) \hat{a}_q^+ \qquad (5.5)$$

The first term in the right member of Eq. (5.5) corresponds to the output field from the modulator, while the second term identifies the radiated field. Turning the attention to the output field term, it is observed that the real part of $S_{q,n}$ is zero if $q-n$ is even. This means that no modes corresponding to even harmonics are expected at the modulator's output. This is a well-known characteristic of this particular modulator design under classical operation.

The second interesting operation regime is the *Single Sideband (SSB) operation* [12] in which by properly dephasing by 90º one of the input RF modulating tones one of the RF sidebands (upper or lower) is eliminated at the output of the modulator. This operation mode is also of widespread use in analog communication due to its resilience to link's dispersion, and represents a scattering device that maximizes the transition probability to a single (upper or lower) sideband mode in one of the two output ports. In this case the values of the parameters are $m_1 = m_2 = m$, $\theta_1 = 0$, $\theta_2 = \pi/2$, $\varphi_{B1} = \pi/2$ and $\varphi_{B2} = 0$, which substituted in Eq. (3.15) yields:

$$S_{q,n} = i\left[(-i)^{q-n} J_{q-n}(m) - (-i)^{q+n} J_{q+n}(m)\right]$$
$$\overline{S}_{q,n} = i^{q-n}\left[(-i)^{q-n} J_{q-n}(m) - (-i)^{q+n} J_{q+n}(m)\right] = i^{q-n-1} S_{q,n} \quad (5.6)$$

Note that for the special case when $q = n-1$, that is, for the lower RF sideband:

$$\overline{S}_{n-1,n} = -S_{n-1,n} \quad (5.7)$$

and thus, as expected, the contribution to the lower RF sideband is cancelled in the first term of the right hand-side member of Eq. (5.3). It can be readily checked that the upper sideband is cancelled if $\theta_1 = 0$, $\theta_2 = -\pi/2$ while keeping unaltered the rest of the parameters.

## 5.3. Low-index multi-tone and cascaded single-tone modulation

In classical terms, multi-tone phase or amplitude modulation has applications such as the quantification of the nonlinear distortions imparted by modulation or the analysis of subcarrier multiplexed systems, where different microwave frequencies $\Omega_k$ are data-bearing carriers of various electrical transmission channels. From the quantum side, and generalizing the results in the previous example, the most immediate application of multi-tone modulation is the operation of modulators as electrically-controlled single or dual-output frequency splitters where the output frequencies are now of the form $\omega_0 + m\Omega_1 + n\Omega_2 + \ldots$ Unfortunately, its analysis is more difficult than for single-tone modulation. The reasons for the complexity of the quantum operation are twofold. First, the intrinsic nonlinearity of the devices, which creates sideband frequencies from single-mode inputs, also originates distortion yet at classical level. And secondly, modulation with different tones is not the concatenation or cascading of the modulation associated to each tone, since the corresponding operators do not commute.

In practical operation conditions, the modulation indices $m_k = mA_k$ associated to different tones or subcarriers in Eq. (3.32) are $\ll 1$, so that an approximation based on the first-order perturbative expansion is justified. Here phase modulation will be discussed; the modulation in amplitude has been analyzed with more detail in [65]. Equation (3.32) implies that, to first order in modulation index, the combined phase-modulation operator can be decomposed in single-tone modulations. For instance, for two-tone modulation,

$$\hat{S}_{PM} = \hat{S}_{PM,2}\hat{S}_{PM,1} + \hat{O}(m^2) = \hat{S}_{PM,1}\hat{S}_{PM,2} + \hat{O}(m^2) \quad (5.8)$$

where $\hat{O}(m^2)$ represents an error operator term of the order $m^2$ and $\hat{S}_{PM,k}$ ($k = 1,2$) are single-tone phase modulation operators, and therefore this approximation reduces the problem to a tractable form. For instance, in FC [25, 29] or SCM-QKD [30] systems the input are in coherent states, so that the analogy with the classical calculus is more apparent. Using Eq. (3.18) with the approximation $S_{k,q} \cong C_{k-q}$ yields:

$$\hat{S}_{PM}\hat{D}_{\omega_0}(\alpha)\hat{S}_{PM}^+ \cong \hat{S}_{PM,2}\hat{S}_{PM,1}\hat{D}_{\omega_0}(\alpha)\hat{S}_{PM,1}^+\hat{S}_{PM,2}^+ = \hat{S}_{PM,2} \bigotimes_{q=q_{\min}}^{\infty} \hat{D}_{\omega_0 + q\Omega_1}(C_q\alpha)\hat{S}_{PM,2}^+ \quad (5.9)$$

where the input radiation is assumed at mode $\omega_0$, $\hat{D}_\omega(\alpha)$ denotes the displacement operator at frequency $\omega$, and $q_{min}$ is the lowest index allowed by positive frequencies. Now, for low modulation indices only the first sidebands are significant since $J_0(m) \cong 1$ and $J_{\pm 1}(m) \cong \pm m/2$ for low $m$, the remaining Bessel functions being of order $m^2$ or higher. Then,

$$C_q = \delta_{q,0} - i\frac{1}{2}m_1 e^{i\theta}\delta_{q,-1} - i\frac{1}{2}m_1 e^{-i\theta}\delta_{q,1} + O(m^2) \quad (5.10)$$

and only two sidebands are generated, while the carrier is left invariant:

$$\hat{S}_{PM,1}\hat{D}_{\omega_0}(\alpha)\hat{S}_{PM,1}^+ \cong \hat{D}_{\omega_0 - \Omega_1}(-ie^{i\theta_1}m_1\alpha/2)\hat{D}_{\omega_0}(\alpha)\hat{D}_{\omega_0 + \Omega_1}(-ie^{-i\theta_1}m_1\alpha/2) \quad (5.11)$$

For the second tone the process is repeated,

$$\hat{S}_{PM}\hat{D}_{\omega_0}(\alpha)\hat{S}_{PM}^+ \cong \hat{S}_{PM,2}\hat{S}_{PM,1}\hat{D}_{\omega_0}(\alpha)\hat{S}_{PM,1}^+\hat{S}_{PM,2}^+ =$$
$$\hat{D}_{\omega_0 - \Omega_2}(-ie^{i\theta_2}m_2\alpha/2)\hat{D}_{\omega_0 - \Omega_1}(-ie^{i\theta_1}m_1\alpha/2)\hat{D}_{\omega_0}(\alpha)\hat{D}_{\omega_0 + \Omega_1}(-ie^{-i\theta_1}m_1\alpha/2)\hat{D}_{\omega_0 + \Omega_2}(-ie^{-i\theta_2}m_2\alpha/2) \quad (5.12)$$

and two additional sidebands are generated from the carrier as in the classical operation.

A similar computation is central in the analysis of FC-QKD systems where a cascaded modulation over the same microwave tone $\Omega_0$ is used: the first low-index modulation is used by Alice to encoded qubits in the phases of the first sidebands, whereas Bob uses a second phase modulator to generate additional sidebands at the same frequency that produce interference at $\omega_0 \pm \Omega_0$. In this case, the separation $\hat{S}_{PM} = \hat{S}_{PM,B}\hat{S}_{PM,A}$ is exact, as the modulators are acting independently at different locations. Then, use of the properties of composition of displacement operators and Eq. (5.12) yields:

$$\hat{S}_{PM,B}\hat{S}_{PM,A}\hat{D}_{\omega_0}(\alpha)\hat{S}^+_{PM,A}\hat{S}^+_{PM,B} \cong$$
$$\hat{D}_{\omega_0-\Omega_0}[-i(e^{i\theta_A}m_A + e^{i\theta_B}m_B)\alpha/2]\hat{D}_{\omega_0}(\alpha)\hat{D}_{\omega_0+\Omega_0}[-i(e^{-i\theta_A}m_A + e^{-i\theta_B}m_B)\alpha/2] \quad (5.13)$$

thus obtaining the desired interference at each sideband mode.

*5.4. Frequency-Coded Quantum Key Distribution*

The previous development will be used here to analyze with mode detail FC QKD systems, which aim at implementing discrete-variables protocols by encoding qubits in the spectral sidebands originated by modulation. In the first reported system the protocol implemented was B92 [23-25]. Here, Alice uses a coherent state a certain frequency $\omega$ which is phase-modulated with low modulation index, $m_A \ll 1$, so that only the sidebands at frequencies $\omega \pm \Omega$ are generated as shown in (5.11). She encodes the qubits by choosing radio-frequency phases $\theta_1$ in the set $\{0, \pi\}$, whereas Bob uses a second phase modulator to generate additional sidebands from the carrier $\omega$ at the same frequency $\Omega$, with phases $\theta_2$ chosen again in the set $\{0, \pi\}$. These new sidebands produce interference at $\omega \pm \Omega$ as shown in (5.13). Bob's modulation index coincides with that of Alice, so that the interference produced at each sideband is constructive only if $\theta_1$ and $\theta_2$ coincide, otherwise the interference nullifies the state at both sidebands. Bob measures only one sideband, say that at $\omega+\Omega$ and, since these sidebands carry weak coherent states, this interference can be regarded as single photon interference between the quantum fields at the sideband generated form the carrier by Alice and Bob. The absence of detection of a photon at this sideband can be due to an incorrect choice of basis or to imperfect detection, so it is useless for sharing a key. But when Bob detects a photocount he infers that his choice of basis was correct. After public announcement, this permits Alice and Bob to share a bit encoded in the basis chosen.

This idea was subsequently modified to implement the BB84 protocol by use of amplitude modulators and detection in two sidebands [26, 27] and also of a combination of amplitude and phase modulators [28]. In all these systems, the qubits are not strictly encoded in frequency, as the name Frequency-Coded may suggest, but in the relative phase between carrier and sidebands. Therefore, these original systems are translations to the frequency domain of traditional phase-coded setups. A truly Frequency-Coded QKD system has been recently demonstrated [29]. Here, phase modulation is used to implement BB84 from four different states, forming a pair of nonconjugate basis in a two-dimensional subspace of the three-dimensional space of one-photon states with frequencies $\omega$ and $\omega \pm \Omega$.

The system is built from a phase-modulated single-photon source at frequency $\omega$. The phase-modulated output is subsequently filtered to the first sidebands so that, according to Eq. (3.22), it can be written, up to a global phase, as the following vector:

$$\hat{S}_{PM}|1\rangle_\omega = \frac{1}{\sqrt{\eta}}\left[J_0(m)|1\rangle_\omega - ie^{-i\theta}J_1(m)|1\rangle_{\omega+\Omega} + ie^{i\theta}J_{-1}(m)|1\rangle_{\omega-\Omega}\right] \quad (5.14)$$

The normalization $\eta$ also describes the efficiency of the generation, i.e., the probability that the phase-modulated photons are detected either in the carrier or in the first sidebands, and is given by $\eta(m) = J_0(m)^2 + 2J_1(m)^2$. Now, Alice can generate the following four states:

$$|\pm,1\rangle = \frac{1}{\sqrt{2}}|1\rangle_\omega \pm \frac{1}{2}|1\rangle_{\omega+\Omega} \mp \frac{1}{2}|1\rangle_{\omega-\Omega} \qquad |+,2\rangle = |1\rangle_\omega \qquad |-,2\rangle = \frac{1}{\sqrt{2}}|1\rangle_{\omega+\Omega} - \frac{1}{\sqrt{2}}|1\rangle_{\omega-\Omega} \quad (5.15)$$

for specific choices of modulation indices and modulation phases: $m = 1.161$ and $\theta = \mp \pi/2$ for the first two states, $m = 0$ for the third, and $m = 2.405$, $\theta = \pi/2$ for the last. These four states represents two nonconjugate basis of the two-dimensional space spanned by vectors $|1\rangle_\omega$ and $|1\rangle_{\omega-\Omega} - |1\rangle_{\omega+\Omega}$ and can therefore be used for implementing the BB84 protocol. The sates in Eq. (5.15) are not produced with the same probability, but this unbalance can be compensated for by use of an uneven generation rate.

Bob's detection is performed with a second phase modulator followed by two detectors, $D_2$ for the carrier frequency and $D_1$ for all the other frequencies in both upper and lower sidebands. If Bob assumes that the basis $|\pm,2\rangle$ has been sent, he produces no phase modulation, so that $D_2$ fires for $|+,2\rangle$ and $D_1$ fires for $|-,2\rangle$ with unit probability. For the basis $|\pm,1\rangle$, he impinges again a phase modulation with $m = 1.161$ and $\theta = -\pi/2$, which implies that $D_1$ fires for $|+,1\rangle$ with unit probability. In contrast, for $|-,1\rangle$ $D_2$ fires with probability 0.953 and thus $D_1$ fires with probability 0.047. This imperfect detection scheme of $|-,1\rangle$ causes an additional decrease in QBER below 1.2% [29].

*5.5. DC Electro-optic Amplitude Modulator with two equal sinusoidal modulating inputs: Switching between two port entangled and separable states.*

The examples presented in section 5.2 have illustrated the multimode transition capabilities over one-photon or coherent states allowed by conventional designs of EOM subject to sinusoidal modulating inputs. However, as interferometric devices with control of the relative phase, modulation index and modulating driving voltage in each arm, EOMs permit compact implementations of standard, single-mode, quantum effects. In fact, if the modulation indices of the phase modulators are set to zero, they behave as phase shifters after the control of the bias voltage. Then, tasks based on bulk optics interferometers can be directly translated to the DC modulators of type C) and D) in figure 4.2, since the four ports of the interferometer are physically accessible and the mathematical description of beamsplitter and directional coupler are the same. However, proper selection and operation of the EOM permit the integration in the same device of phase modulator and interferometer, thus allowing a direct extension of certain operations to multimode, phase-modulated fields. In this subsection it is analyzed the use of DC modulators for switching and interpolating between two-port entangled and separable states composed of phase-modulated, multimode photons. Let us assume a two-photon input (one at each input port) in the same mode under single-tone driving and use the notations in Eq. (5.2). The coefficients of DC modulators are:

$$t_o = t_o^{'} = t_i = t_i^{'} = 1/\sqrt{2} \qquad r_i^{'} = r_i = r_o = r_o^{'} = i/\sqrt{2} \qquad (5.16)$$

Then, use of (4.6) and (4.7) yields:

$$\hat{S}_{EOM}\left(\hat{a}_n^+ \otimes \hat{a}_n^+\right)\hat{S}_{EOM}^+ = \hat{S}_{EOM}\left(\hat{a}_n^+ \otimes \hat{1}\right)\hat{S}_{EOM}^+ \hat{S}_{EOM}\left(\hat{1} \otimes \hat{a}_n^+\right)\hat{S}_{EOM}^+ =$$
$$= \frac{i}{4}\left(\hat{b}_n^+\hat{b}_n^+ \otimes \hat{1}\right) - \frac{1}{2}\left(\hat{b}_n^+ \otimes \hat{b}_n^+\right) - \frac{i}{4}\left(\hat{1} \otimes \hat{b}_n^+\hat{b}_n^+\right) - \frac{i}{4}\left(\hat{c}_n^+\hat{c}_n^+ \otimes \hat{1}\right) - \frac{1}{2}\left(\hat{c}_n^+ \otimes \hat{c}_n^+\right) + \frac{i}{4}\left(\hat{1} \otimes \hat{c}_n^+\hat{c}_n^+\right) \qquad (5.17)$$

It is observed that the two photons circulate in the same modulator's arm because both emerge together after the input beamsplitter [57-61]. Furthermore, if equal modulation indices and radio-frequency driving tones are set in both phase modulators, the phase-modulation operators are related by $\hat{c}_n^+ = \exp(i\Delta\varphi_B)\hat{b}_n^+$ as can be derived from Eq. (5.2), where $\Delta\varphi_B$ represents the phase difference induced by different bias voltages between the upper and lower phase modulators. Hence, Eq. (5.17) can be expressed as:

$$\hat{S}_{EOM}\left(\hat{a}_n^+ \otimes \hat{a}_n^+\right)\hat{S}_{EOM}^+ = e^{i\Delta\varphi_B}\sin(\Delta\varphi_B)\frac{1}{2}[\hat{b}_n^+\hat{b}_n^+ \otimes \hat{1} - \hat{1} \otimes \hat{b}_n^+\hat{b}_n^+] - e^{i\Delta\varphi_B}\cos(\Delta\varphi_B)\,\hat{b}_n^+ \otimes \hat{b}_n^+ \qquad (5.18)$$

Now, the structure of the output state can be controlled by $\Delta\varphi_b$. For instance, if $\Delta\varphi_b = 0$ then the output is a product state, whereas for $\Delta\varphi_b = \pi/2$ the output state is entangled, and at this point the overall modulator behaves in a similar fashion to a balanced beamsplitter acting on phase-modulated states.

*5.6. Single-tone amplitude modulation of a single photon wavepacket*

In the following sections the amplitude modulation of the wavepackets describing single-photon states will be addressed [49]. Let us first consider a single-photon wavepacket at the input of a Y-branch modulator sinusoidally modulated at frequency $\Omega_0$. The wavepacket is described by a frequency spread $\varphi(\omega)$, which is assumed narrowband and centered at optical frequency $\omega_0$, or by its wavepacket amplitude $\varphi(t)$. These functions are normalized such that $\int|\varphi(\omega)|^2 d\omega/2\pi = \int|\varphi(t)|^2 dt = 1$, and the quantum states is [57]:

$$\left|1_\varphi\right\rangle = \int_0^\infty \frac{d\omega}{2\pi}\varphi(\omega)\hat{a}^+(\omega)\left|vac\right\rangle = \int_{-\infty}^\infty dt\varphi(t)\hat{a}^+(t)\left|vac\right\rangle \qquad (5.19)$$

Due to the narrowband approximation, integrals in $\omega$ can be extended to $-\infty$ as explained in section 3.3. The beamsplitter coefficient specialization is that of Eq. (5.1), and the action of the EOM can be described using Eq. (4.6) with single-tone PM in continuous-mode formalism, which reads:

$$\hat{b}^+(\omega) \equiv \hat{S}_{PM_1}\hat{a}^+(\omega)\hat{S}^\dagger_{PM_1} = \sum_{s=-\infty}^{\infty} C_s\, \hat{a}^+(\omega + s\,\Omega_0)$$
$$\hat{c}^+(\omega) \equiv \hat{S}_{PM_2}\hat{a}^+(\omega)\hat{S}^\dagger_{PM_2} = \sum_{s=-\infty}^{\infty} \bar{C}_s\, \hat{a}^+(\omega + s\,\Omega_0)$$
(5.20)

after the approximation in Eq. (3.22) where the scattering coefficients $S_{k,q}$ are substituted by their classical values, $S_{k,q} \cong C_{k-q}$ and the extension of the sums up to $s = -\infty$ assuming that the coefficients allowing for transitions to negative frequencies are negligible. Hence,

$$\hat{S}_{EOM}\left|1_\varphi\right\rangle \otimes \left|vac\right\rangle = \frac{1}{2}\left[\sum_{s=-\infty}^{\infty}(C_s + \bar{C}_s)\int_0^\infty \frac{d\omega}{2\pi}\varphi(\omega)\hat{a}^+(\omega+s\Omega_0)\left|vac\right\rangle \otimes \left|vac\right\rangle\right] + $$
$$+ \frac{1}{2}\left[\left|vac\right\rangle \otimes \sum_{s=-\infty}^{\infty}(-C_s + \bar{C}_s)\int_0^\infty \frac{d\omega}{2\pi}\varphi(\omega)\hat{a}^+(\omega+s\Omega_0)\left|vac\right\rangle\right]$$
(5.21)

and, after the change of variables $\omega' = \omega + s\,\Omega_0$ and the extension of the lower limits of the integrals in $\omega'$,

$$\hat{S}_{EOM}\left|1_\varphi\right\rangle \otimes \left|vac\right\rangle = \int_0^\infty \frac{d\omega'}{2\pi}\varphi_O(\omega')\hat{a}^+(\omega')\left|vac\right\rangle \otimes \left|vac\right\rangle + \left|vac\right\rangle \otimes \int_0^\infty \frac{d\omega'}{2\pi}\varphi_R(\omega')\hat{a}^+(\omega')\left|vac\right\rangle$$
$$= \left|1_{\varphi_O}\right\rangle \otimes \left|vac\right\rangle + \left|vac\right\rangle \otimes \left|1_{\varphi_R}\right\rangle$$
(5.22)

where $\varphi_O(\omega)$ and $\varphi_R(\omega)$ are the resulting (non-normalized) wavepackets describing the amplitude modulated photon corresponding to the output and radiated fields, respectively:

$$\varphi_O(\omega) = \frac{1}{2}\sum_{s=-\infty}^{\infty}(C_s + \bar{C}_s)\varphi(\omega - s\Omega_0) \qquad \varphi_R(\omega) = \frac{1}{2}\sum_{s=-\infty}^{\infty}(-C_s + \bar{C}_s)\varphi(\omega - s\Omega_0)$$
(5.23)

These equations admit a neat description in the time domain. From Eq. (5.23), the action of the amplitude modulator in terms of wavepacket amplitudes is multiplicative,

$$\varphi_O(t) = m_O(t)\varphi(t) \qquad \varphi_R(t) = m_R(t)\varphi(t)$$
(5.24)

where

$$m_O(t) = m_{11}(t) = \frac{1}{2}\sum_{s=-\infty}^{\infty}(C_s + \bar{C}_s)\exp(-is\Omega_0 t) = \frac{1}{2}\left[e^{i\alpha_1(t)} + e^{i\alpha_2(t)}\right]$$
$$m_R(t) = m_{21}(t) = \frac{1}{2}\sum_{s=-\infty}^{\infty}(-C_s + \bar{C}_s)\exp(-is\Omega_0 t) = \frac{1}{2}\left[-e^{i\alpha_1(t)} + e^{i\alpha_2(t)}\right]$$
(5.25)

are the modulation functions of the classical description of the modulator for the output and radiated ports, respectively, and $\alpha_k(t) = \varphi_{Bk} - m_k\cos(\Omega_0 t + \theta_k)$ is the phase impinged by each modulator. Clearly, the wavepacket amplitudes $\varphi_O$ and $\varphi_R$ in Eq. (5.24) are not normalized, but despite the approximations involved the probabilistic interpretation of the output state holds since:

$$\int dt\,(|\varphi_O(t)|^2 + |\varphi_R(t)|^2) = \int dt\,|\varphi(t)|^2(|m_O(t)|^2 + |m_R(t)|^2) = \int dt\,|\varphi(t)|^2 = 1$$
(5.26)

which simply reflects that the photon exits either through the physical output port or through the radiation port. The normalized states $\left|1_{\varphi O}\right\rangle' = \left|1_{\varphi O}\right\rangle / \left\langle 1_{\varphi O}|1_{\varphi O}\right\rangle^{1/2}$ and $\left|1_{\varphi R}\right\rangle' = \left|1_{\varphi R}\right\rangle / \left\langle 1_{\varphi R}|1_{\varphi R}\right\rangle^{1/2}$ describe the conditional output states resulting from the actual observation of the photon at the physical output port or at the radiation port, respectively. The generalization of this result to arbitrary phase or amplitude modulation is straightforward: within the approximation of narrowband optical inputs, the wavepacket amplitude is multiplied by the classical modulation functions as in Eq. (5.24).

*5.7. Two-photon interference with phase-modulated inputs*

In this example it will be analyzed the experiment of two-photon interference with a phase-modulated wavepacket reported in [48]. Referring to figure 5.1, each of the input ports of a beamsplitter is fed by a product state composed of two photons from the same ensemble, thus described by the same wavepacket amplitude, $\varphi(t)$. One of these photons undergoes phase modulation before impinging the first input port of the beamsplitter. Its wavepacket is $\bar{\varphi}(t) = \varphi(t)\exp[i\alpha(t)]$ with $\alpha(t)$ the classical driving function. Then, the out state is:

$$|out\rangle = \hat{S}_{BS}(\hat{S}_{PM} \otimes \hat{1})|1_\varphi\rangle \otimes |1_\varphi\rangle = \hat{S}_{BS}|1_{\bar\varphi}\rangle \otimes |1_\varphi\rangle =$$

$$= \frac{i}{2}\int_{-\infty}^{\infty} dt \int_{-\infty}^{\infty} dt' \, \bar\varphi(t)\varphi(t')[\hat{a}^+(t)\hat{a}^+(t') \otimes \hat{1} + \hat{1} \otimes \hat{a}^+(t)\hat{a}^+(t')]|vac\rangle \otimes |vac\rangle \qquad (5.27)$$

$$+ \frac{1}{2}\int_{-\infty}^{\infty} dt \int_{-\infty}^{\infty} dt' \, [\bar\varphi(t)\varphi(t') - \varphi(t)\bar\varphi(t')][\hat{a}^+(t) \otimes \hat{a}^+(t')]|vac\rangle \otimes |vac\rangle$$

The different probabilities associated to a coincidence-counting experiment can be extracted from the following resolution of the identity of the two-photon, continuous-mode Fock space:

$$\hat{1}_{\mathcal{F}_2} = \frac{1}{2}\int_{-\infty}^{\infty} dt_1 dt_2 \, [\hat{a}^+(t_1)\hat{a}^+(t_2)\hat{a}(t_2)\hat{a}(t_1) \otimes \hat{1} + \hat{1} \otimes \hat{a}^+(t_1)\hat{a}^+(t_2)\hat{a}(t_2)\hat{a}(t_1)]$$

$$+ \int_{-\infty}^{\infty} dt_1 dt_2 \, [\hat{a}^+(t_1)\hat{a}(t_1) \otimes \hat{a}^+(t_2)\hat{a}(t_2)] \qquad (5.28)$$

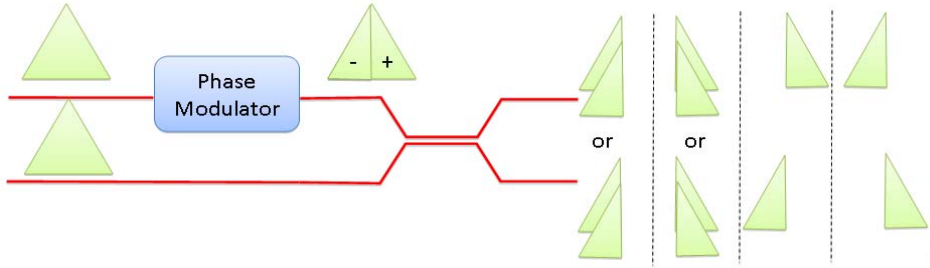

**Figure 5.1.** Two-photon interference with phase-modulated (PM) photons: the temporal wavepacket spread is represented by a triangle. After modulation of one of the photons with a π phase shift they are mixed in a beamsplitter. The resulting coincidences show same-port coalescence or rejection depending on the detection instants, which are controlled by the phase modulation function.

From this it is straightforward to extract the probability $P(mt_1, nt_2)$ of detecting a photon at port $m$ at time $t_1$ and a photon at port $n$ and at time $t_2$:

$$P(1t_1, 1t_2) = \frac{1}{2}\langle out|\hat{a}^+(t_1)\hat{a}^+(t_2)\hat{a}(t_2)\hat{a}(t_1) \otimes \hat{1}|out\rangle$$

$$= \frac{1}{4}|\varphi(t_1)|^2|\varphi(t_2)|^2\left[1 + \cos(\alpha(t_1) - \alpha(t_2))\right] = P(2t_1, 2t_2)$$

$$P(1t_1, 2t_2) = \frac{1}{2}\langle out|\hat{a}^+(t_1)\hat{a}(t_1) \otimes \hat{a}^+(t_2)\hat{a}(t_2)|out\rangle = \qquad (5.29)$$

$$= \frac{1}{2}|\varphi(t_1)|^2|\varphi(t_2)|^2\left[1 - \cos(\alpha(t_1) - \alpha(t_2))\right]$$

Then, both photons emerge from the same output port of the beamsplitter if the phase modulation is null ($\alpha$=0): this is the usual spatial 'bosonic' behavior. But if the impinged phase modulation varies along the photon temporal spread, there is a finite probability of detecting the photons in opposite ports as it depends on the difference in phase shift between detection times. In [48], wavepackets with long temporal spread were used, and the modulation was zero for the photon's leading edge and π for the trailing edge. Coincidences in the same output port were found when the observation times correspond to either the leading or the trailing edges otherwise the coincidences appear between opposite ports. This behavior is schematically depicted in figure 5.1, and can be viewed as a generalization of the example in section 5.5 to wavepacket level.

### 5.8. Amplitude modulation of single photons as a quantum operation

The electro-optic amplitude modulator, as described by Eq. (4.4), is a 2×2 device where, depending on the particular geometry chosen, only one or both input or output ports are physically accessible. For instance, for Y-Branch modulators there is only a single physical input and output ports accessible, which means that the radiation input port is always set to the vacuum state and the radiation output port is not disposable for any quantum task. Then, it is natural to ask for the description of quantum modulators where a general single input port state is modulated and only a single output port state is measured or used in a subsequent processing step. This leads to a formulation based on a (trace-preserving) quantum operation [77],

$$\hat{\rho}_{in} \to \hat{\rho}_{out} = T(\hat{\rho}_{in}) = tr_2[\hat{S}(\hat{\rho}_{in} \otimes |vac\rangle\langle vac|)\hat{S}^+] \tag{5.30}$$

where $\hat{\rho}_{in}$ and $\hat{\rho}_{out}$ are the density matrices describing one port input and output states, $\hat{S}$ represents a unitary 2×2 evolution, which in this case is that of amplitude modulator $\hat{S} = \hat{S}_{EOM}$, $tr_2$ traces out the output radiation port and $T$ represents the quantum operation at the level of density matrices.

In the present case the unitary evolution conserves the number of photons, $[\hat{N}, \hat{S}_{EOM}] = 0$. This condition restricts the form of the output, one-port state, when the input state has a definite number of photons, say $n$. Denoting by $\hat{\rho}_{in}^{(n)}$ such an input, the form of the output must be a convex combination of states with a definite number of photons less than or equal to $n$:

$$\hat{\rho}_{in}^{(n)} \to \hat{\rho}_{out} = T(\hat{\rho}_{in}) = \sum_{k=0}^{n} p_k \hat{\rho}_{out}^{(k)} \tag{5.31}$$

with $\hat{\rho}_{out}^{(k)} = \hat{\rho}_{out}^{(k)+} \geq 0$, $tr(\hat{\rho}_{out}^{(k)}) = 1$, $p_k \geq 0$ and $\sum_{k=0}^{n} p_k = 1$. The proof of this result is given in the appendix. Essentially, this means that the output corresponding to an input with $n$ photons is in general a mixed state that can carry up to $n$ photons. The density matrices $\hat{\rho}_{out}^{(k)}$ represent the output state conditioned to the actual exit of exactly $k$ photons, a fact that occurs with probability $p_k$.

If the input contains just one photon, the form given by Eq. (5.31) is particularly simple,

$$\hat{\rho}_{in}^{(1)} \to \hat{\rho}_{out} = T(\hat{\rho}_{in}) = p_0 |vac\rangle\langle vac| + p_1 \hat{\rho}_{out}^{(1)} \tag{5.32}$$

and can be easily worked: for an arbitrary pure state input $|\Psi\rangle$ containing just one photon in port 1 (input port 2 is in vacuum state) the output is, according to Eq. (4.5),

$$\hat{S}_{EOM}(|\Psi\rangle \otimes |vac\rangle) = (\hat{V}|\Psi\rangle) \otimes |vac\rangle + |vac\rangle \otimes (\hat{W}|\Psi\rangle) \tag{5.33}$$

where $\hat{V}$ and $\hat{W}$ stand for operators $\hat{V} = t_i' t_o \hat{S}_{PM_1} + r_i' r_o \hat{S}_{PM_2}$ and $\hat{W} = t_i' r_o \hat{S}_{PM_1} + r_i' t_o \hat{S}_{PM_2}$. From the unitarity of $\hat{S}_{EOM}$ it can be shown that these operators verify $\hat{V}^+\hat{V} + \hat{W}^+\hat{W} = \hat{1}$. The pure-state relation in Eq. (5.33) can be easily extended to mixed states $\hat{\rho}_{in}^{(1)} = \sum_k \lambda_k |\Psi_k\rangle\langle\Psi_k|$:

$$\begin{aligned}\hat{S}_{EOM}(\hat{\rho}_{in}^{(1)} \otimes |vac\rangle\langle vac|)\hat{S}_{EOM}^+ &= (\hat{V}\hat{\rho}_{in}^{(1)}\hat{V}^+) \otimes |vac\rangle\langle vac| + |vac\rangle\langle vac| \otimes (\hat{W}\hat{\rho}_{in}^{(1)}\hat{W}^+) \\ &+ \sum_k \lambda_k (\hat{V}|\Psi_k\rangle\langle vac| \otimes |vac\rangle\langle\Psi_k|\hat{W}^+ + |vac\rangle\langle\Psi_k|\hat{V}^+ \otimes \hat{W}|\Psi_k\rangle\langle vac|)\end{aligned} \tag{5.34}$$

Inserting now this expression in Eq. (5.30) and computing the trace, it is observed that the interference terms in the second line of Eq. (5.34) vanish, since $\hat{W}|\Psi_k\rangle$ and $\hat{V}|\Psi_k\rangle$ are one-photon states and thus orthogonal to the vacuum. Then, comparison with Eq. (5.32) yields:

$$p_0 = tr[\hat{W}\hat{\rho}_{in}^{(1)}\hat{W}^+] \qquad p_1 = tr[\hat{V}\hat{\rho}_{in}^{(1)}\hat{V}^+] \qquad \hat{\rho}_{out}^{(1)} = \hat{V}\hat{\rho}_{in}^{(1)}\hat{V}^+ / tr[\hat{V}\hat{\rho}_{in}^{(1)}\hat{V}^+] \tag{5.35}$$

It is apparent that $p_k \geq 0$ and it can also be easily verified that the sum over probabilities equals unity, $p_0 + p_1 = 1$, as a consequence of the unitarity of $\hat{S}_{EOM}$. Considering as a particular example a pure input state described by the wavepacket in Eq. (5.19), it is immediate to show, using the same approximations as in section 5.6, that $\hat{\rho}_{out}^{(1)}$ indeed describes a pure one-photon state with normalized wavepacket amplitude $\varphi_O(t) = m_O(t)\varphi(t) \times (\int |m_O(t)\varphi(t)|^2 dt)^{-1/2}$, where:

$$m_O(t) = m_{11}(t) = t_i' t_o e^{i\varphi_{B1} - im_1 x_1(t)} + r_i' t_o e^{i\varphi_{B2} - im_2 x_2(t)} \tag{5.36}$$

is the classical modulation function of the modulator's physical output (output port 1) when the input is also in port one. Finally, the quantum operation over one-photon states can be easily described in the operator-sum representation:

$$\hat{\rho}_{out} = T(\hat{\rho}_{in}^{(1)}) = \hat{K}_0 \hat{\rho}_{in}^{(1)} \hat{K}_0^+ + \int_0^\infty \frac{d\omega}{2\pi} \hat{K}(\omega) \hat{\rho}_{in}^{(1)} \hat{K}(\omega)^+ \tag{5.37}$$

where the Kraus operators are $\hat{K}_0 = \hat{V}$ and:

$$\hat{K}(\omega) = |vac\rangle\langle 1_\omega|\hat{W} = |vac\rangle\langle vac|\hat{a}(\omega)\hat{W} \tag{5.38}$$

*5.9. Correlations of phase and amplitude modulated states*

In this final example the quantum correlations after general amplitude or phase modulation will be addressed. The analysis is based on the formalism of the complex envelope, which is particularly useful to highlight the similarities with the classical counterpart, and is restricted to first-order correlations since the extension to second-order correlations, which are the basic quantities in the analysis of photon-counting experiments [50], is straightforward. The correlator of an input state is given by

$$G_{in}^{(1)}(t_1,t_2) = \langle \hat{E}^{(-)}(t_1)\hat{E}^{(+)}(t_2) \rangle \propto \langle \hat{A}^+(t_1)\hat{A}(t_2) \rangle \exp[-i\omega_0(t_2-t_1)] \qquad (5.39)$$

where the brackets denote quantum average over the state, $\langle \hat{O} \rangle = tr[\hat{\rho}_{in}\hat{O}]$ and using Eq. (3.35) the correlator has been rewritten in terms of the complex envelope, assuming that the last equality only holds under the hypothesis in section 3.3: that the spectral content of the relevant quantum state $\hat{\rho}_{in}$ is narrowband and belongs to the optical band. Then, the change experienced by the quantum state after phase modulation is, in the Schrödinger image $\hat{\rho}_{in} \to \hat{\rho}_{out} = \hat{S}_{PM}\hat{\rho}_{in}\hat{S}_{PM}^+$, which can be brought to the Heisenberg image as a change in $\hat{A}(t)$ given by $\hat{A}(t) \to \hat{S}_{PM}^+\hat{A}(t)\hat{S}_{PM}$. Then, Eq. (3.39) implies that:

$$G_{PM}^{(1)}(t_1,t_2) = G_{in}^{(1)}(t_1,t_2)\exp\{im[x(t_1)-x(t_2)]\} \qquad (5.40)$$

Therefore, and with the narrowband approximation, the change in the correlation function induced by phase modulation coincides with the multiplicative change induced by phase modulation in the classical correlations. In addition, and in a similar fashion to the classical theory, the state after phase modulation is in general not stationary, even if the input state was.

As for the amplitude modulator the situation is similar, with the only complications derived from the existence of dual inputs and outputs. For instance, the simplest case is that of an input state $\hat{\rho}$ in a single input port, say 1: $\hat{\rho} = \hat{\rho}_{in} \otimes |vac\rangle\langle vac|$, and correlations can be measured in the same output port or between different output ports. These correlations are thus proportional to:

$$G_{EOM,mn}^{(1)}(t_1,t_2) = \langle \hat{E}_m^{(-)}(t_1)\hat{E}_n^{(+)}(t_2) \rangle \qquad (5.41)$$

where $m, n = 1, 2$, $\hat{E}_1^{(\pm)}(t) = \hat{E}^{(\pm)}(t) \otimes \hat{1}$ and $\hat{E}_2^{(\pm)}(t) = \hat{1} \otimes \hat{E}^{(\pm)}(t)$. Let us consider $G_{EOM,12}^{(1)}$, the other calculations are similar. The computation can be brought to Heisenberg image as in Eq. (5.40):

$$\begin{aligned}\langle \hat{A}^+(t_1) \otimes \hat{A}(t_2) \rangle &= tr_{12}[\hat{S}_{EOM}(\hat{\rho}_{in} \otimes |vac\rangle\langle vac|)\hat{S}_{EOM}^+(\hat{A}^+(t_1) \otimes \hat{A}(t_2))] \\ &= tr_{12}[(\hat{\rho}_{in} \otimes |vac\rangle\langle vac|)\hat{S}_{EOM}^+(\hat{A}^+(t_1) \otimes \hat{1})\hat{S}_{EOM}\hat{S}_{EOM}^+(\hat{1} \otimes \hat{A}(t_2))\hat{S}_{EOM}]\end{aligned} \qquad (5.42)$$

where $tr_{12}$ traces both output ports. Using Eq. (4.9) the trace contains four terms; three of which vanish since $tr[\hat{A}(t_2)] = tr[\hat{A}^+(t_1)] = 0$. Then,

$$G_{EOM,12}^{(1)}(t_1,t_2) = m_{11}^*(t_1)m_{21}(t_2)G_{in}^{(1)}(t_1,t_2) \qquad (5.43)$$

and the same conclusion as with phase modulation is obtained: the modification introduced by modulation in the quantum correlations coincides with the classical expression. This result can be easily generalized to correlations between any of the output ports. For instance, for the single-photon input wavepacket in Eq. (5.19) the correlation function before modulation is [57]:

$$\langle \hat{E}^{(-)}(t_1)\hat{E}^{(+)}(t_2) \rangle \propto \langle \hat{A}^+(t_1)\hat{A}(t_2) \rangle e^{-i\omega_0(t_1-t_2)} = \langle 1_\varphi | \hat{A}^+(t_1)\hat{A}(t_2) | 1_\varphi \rangle e^{-i\omega_0(t_1-t_2)} = \varphi^*(t_1)\varphi(t_2) \qquad (5.44)$$

and therefore, after amplitude modulation the quantum degree of first-order coherence between output ports $ab$ is:

$$g_{EOM,ab}^{(1)}(t_1,t_2) = \frac{G_{EOM,ab}^{(1)}(t_1,t_2)}{G_{EOM,ab}^{(1)}(t_1,t_1)^{1/2}G_{EOM,ab}^{(1)}(t_2,t_2)^{1/2}} = \frac{m_{a1}^*(t_1)m_{b1}(t_2)}{|m_{a1}(t_1)m_{b1}(t_2)|}\frac{\varphi^*(t_1)\varphi(t_2)}{|\varphi(t_1)\varphi(t_2)|} \qquad (5.45)$$

## 6. Summary and Conclusions

In this paper the salient features of the scattering theory describing both the operation of the electro-optic phase and amplitude modulators subject to single photon and coherent input states have been reviewed. This subject is timely and of importance in light of the increasing utilization of these devices in the framework of quantum information and quantum communication systems.

This review has also provided a consistent description of modulation devices both in discrete mode and in continuous mode formalism. In addition, the paper has included a tutorial development of the use of these models in selected but yet important applications. By choosing and developing several illustrative examples a wide range of topics have been covered, showing the applicability of the models herein presented.

**Appendix**

In this appendix the result described by Eq. (5.31) will be proven. Let us denote by $\hat{\rho}_{in}^{(n)}$ a (eventually mixed) state with $n$ photons in the first port, $\hat{N}_1\hat{\rho}_{in}^{(n)} = \hat{\rho}_{in}^{(n)}\hat{N}_1 = n\hat{\rho}_{in}^{(n)}$ and assume that the interaction conserves its number, $[\hat{N},\hat{S}] = 0$ with $\hat{N} = \hat{N}_1 \otimes \hat{I} + \hat{I} \otimes \hat{N}_2$. Then,

$$\hat{N}\hat{S}(\hat{\rho}_{in}^{(n)} \otimes |vac\rangle\langle vac|)\hat{S}^+ = (\hat{N}_1 \otimes \hat{I} + \hat{I} \otimes \hat{N}_2)\hat{S}(\hat{\rho}_{in}^{(n)} \otimes |vac\rangle\langle vac|)\hat{S}^+$$
$$= \hat{S}\hat{N}(\hat{\rho}_{in}^{(n)} \otimes |vac\rangle\langle vac|)\hat{S}^+ = \hat{S}(\hat{N}_1 \otimes \hat{I} + \hat{I} \otimes \hat{N}_2)(\hat{\rho}_{in}^{(n)} \otimes |vac\rangle\langle vac|)\hat{S}^+ \quad (A.1)$$
$$= \hat{S}(\hat{N}_1\hat{\rho}_{in}^{(n)} \otimes |vac\rangle\langle vac|)\hat{S}^+ = n\hat{S}(\hat{\rho}_{in}^{(n)} \otimes |vac\rangle\langle vac|)\hat{S}^+$$

Taking traces over space 2 in the first and last parts of the equality:

$$tr_2[\hat{N}\hat{S}(\hat{\rho}_{in}^{(n)} \otimes |vac\rangle\langle vac|)\hat{S}^+] = n\hat{\rho}_{out} \quad (A.2)$$

and also, tracing in 2 the first and second expressions in Eq. (A.1):

$$tr_2[\hat{N}\hat{S}(\hat{\rho}_{in}^{(n)} \otimes |vac\rangle\langle vac|)\hat{S}^+] =$$
$$= \hat{N}_1 tr_2[\hat{S}(\hat{\rho}_{in}^{(n)} \otimes |vac\rangle\langle vac|)\hat{S}^+] + tr_2[(\hat{I} \otimes \hat{N}_2)\hat{S}(\hat{\rho}_{in}^{(n)} \otimes |vac\rangle\langle vac|)\hat{S}^+] \quad (A.3)$$
$$= \hat{N}_1\hat{\rho}_{out} + tr_2[(\hat{I} \otimes \hat{N}_2)\hat{S}(\hat{\rho}_{in}^{(n)} \otimes |vac\rangle\langle vac|)\hat{S}^+]$$

Comparison Eqs. (A.3) and (A.2) yields

$$\hat{N}_1\hat{\rho}_{out} = n\hat{\rho}_{out} - tr_2(\hat{L}) \quad (A.4)$$

where it has been defined

$$\hat{L} = (\hat{I} \otimes \hat{N}_2)\hat{S}(\hat{\rho}_{in}^{(n)} \otimes |vac\rangle\langle vac|)\hat{S}^+ \quad (A.5)$$

Clearly $\hat{L}$ is a positive operator since $(\hat{I} \otimes \hat{N}_2)$ and $(\hat{\rho}_{in}^{(n)} \otimes |vac\rangle\langle vac|)$ are positive and $\hat{S}$ is unitary.

Starting now with $\hat{N}$ applied on the right instead of on the left of the first expression in Eq. (A.1) and repeating the steps, yields:

$$\hat{\rho}_{out}\hat{N}_1 = n\hat{\rho}_{out} - tr_2[\hat{S}(\hat{\rho}_{in}^{(n)} \otimes |vac\rangle\langle vac|)\hat{S}^+(\hat{I} \otimes \hat{N}_2)] = n\hat{\rho}_{out} - tr_2(\hat{L}) \quad (A.6)$$

where $tr_2[(\hat{I} \otimes \hat{A}_2)\hat{B}_{12}] = tr_2[\hat{B}_{12}(\hat{I} \otimes \hat{A}_2)]$ has been used. Then, Eqs. (A.4) and (A.6) imply $[\hat{\rho}_{out}, \hat{N}_1] = 0$ and this means that operator $\hat{\rho}_{out}$ can be decomposed in blocks with definite number of photons,

$$\hat{\rho}_{out} = \sum_{k=0}^{\infty} \hat{\Lambda}^{(k)} \quad (A.7)$$

In Eq. (A.7) the sum is direct, as it is a sum over operators acting onto orthogonal subspaces each having $k$ photons. This implies that each of these operators are hermitian and positive, $\hat{\Lambda}^{(k)} = \hat{\Lambda}^{(k)+} \geq 0$, since $\hat{\rho}_{out}$ is hermitian and positive. Now, from Eqs. (A.7) and (A.4),

$$tr_2(\hat{L}) = \sum_{k=0}^{\infty} (n-k)\hat{\Lambda}^{(k)} \quad (A.8)$$

However, the operator $tr_2(\hat{L})$ is also positive because, being $\hat{L}$ positive and $|\Psi\rangle$ an arbitrary state,

$$\langle\Psi|tr_2(\hat{L})|\Psi\rangle = \sum_k \left(\langle\Psi| \otimes \langle v_k|\right)\hat{L}\left(|v_k\rangle \otimes |\Psi\rangle\right) \geq 0 \quad (A.9)$$

where $|v_k\rangle$ is a basis of the second subspace. Then, the sum in Eq. (A.7) must end in $n$ since $\hat{\Lambda}^{(k)} \geq 0$ and:

$$\hat{\rho}_{out} = \sum_{k=0}^{n} \hat{\Lambda}^{(k)} \quad (A.10)$$

Defining finally $p_k = tr(\hat{\Lambda}^{(k)})$ and $\hat{\rho}_{out}^{(k)} = \hat{\Lambda}^{(k)}/tr(\hat{\Lambda}^{(k)})$ equation (5.31) follows. Notice that $\Sigma_k p_k = 1$ since the output state is normalized, $tr(\hat{\rho}_{out}) = tr_{12}(\hat{S}(\hat{\rho}_{in}^{(n)} \otimes |vac\rangle\langle vac|)\hat{S}^+) = tr(\hat{\rho}_{in}^{(n)}) = 1$.

**Acknowledgements**

The authors acknowledge the financial support of the Spanish Government through Project TEC2008-02606 and Project Quantum Optical Information Technology (QOIT), a CONSOLIDER-INGENIO 2010 Project; and also the Generalitat Valenciana through the PROMETEO research excellency award programme GVA PROMETEO 2008/092.